\documentclass[letterpaper]{article} 
\usepackage{aaai25}  
\usepackage{times}  
\usepackage{helvet}  
\usepackage{courier}  
\usepackage[hyphens]{url}  
\usepackage{graphicx} 
\usepackage{natbib}  
\usepackage{caption} 
\usepackage{algorithm}
\usepackage{algorithmic}

\usepackage{overpic}
\usepackage{multirow}
\usepackage{amssymb}
\usepackage{amsmath}
\usepackage{booktabs}
\usepackage{xcolor}

\newcommand{\SZ}[1]{{\color{black}{#1}}}
\newcommand{\nothing}[1]{}
\definecolor{cyanTop}{cmyk}{1,0,0,0}
\definecolor{magentaTopTwo}{cmyk}{0,1,0,0}

\usepackage{tabularx}

\usepackage{newfloat}
\usepackage{listings}
\DeclareCaptionStyle{ruled}{labelfont=normalfont,labelsep=colon,strut=off} 
\floatstyle{ruled}
\newfloat{listing}{tb}{lst}{}
\floatname{listing}{Listing}
\title{GSDiff: Synthesizing Vector Floorplans via \\ Geometry-enhanced Structural Graph Generation}

\author{
    Sizhe Hu\textsuperscript{\rm 1}\thanks{https://orcid.org/0000-0001-9308-5768},
    Wenming Wu\textsuperscript{\rm 2}\thanks{Corresponding author: wwming@hfut.edu.cn (Wenming Wu)}\thanks{https://orcid.org/0000-0002-0640-8520}, 
    Yuntao Wang\textsuperscript{\rm 3}\thanks{https://orcid.org/0009-0003-2827-5727}, 
    Benzhu Xu\textsuperscript{\rm 4}\thanks{https://orcid.org/0000-0001-7756-0901}, 
    Liping Zheng\textsuperscript{\rm 5}\thanks{https://orcid.org/0000-0001-5071-9628}
}

\affiliations{
    \textsuperscript{\rm 1}Hefei University of Technology \\
    485 Danxia Lu, Shushan District, Hefei City, Anhui Province, 230061, China
}

\usepackage{bibentry}

\begin{document}

\maketitle

\begin{abstract}
Automating architectural floorplan design is vital for housing and interior design, offering a faster, cost-effective alternative to manual sketches by architects. However, existing methods, including rule-based and learning-based approaches, face challenges in design complexity and constrained generation with extensive post-processing, and tend to obvious geometric inconsistencies such as misalignment, overlap, and gaps. In this work, we propose a novel generative framework for vector floorplan design via structural graph generation, called GSDiff, focusing on wall junction generation and wall segment prediction to capture both geometric and semantic aspects of structural graphs. To improve the geometric rationality of generated structural graphs, we propose two innovative geometry enhancement methods. In wall junction generation, we propose a novel alignment loss function to improve geometric consistency. In wall segment prediction, we propose a random self-supervision method to enhance the model’s perception of the overall geometric structure, thereby promoting the generation of reasonable geometric structures. Employing the diffusion model and the Transformer model, as well as the geometry enhancement strategies, our framework can generate wall junctions, wall segments and room polygons with structural and semantic information, resulting in structural graphs that accurately represent floorplans. Extensive experiments show that the proposed method surpasses existing techniques, enabling free generation and constrained generation, marking a shift towards structure generation in architectural design. Code and data are available at \url{https://github.com/SizheHu/GSDiff}.
\end{abstract}

%

\section{Introduction} \label{sec:introduction}

Automatic design of architectural floorplans has garnered widespread attention, as detailed architectural blueprints are crucial for constructing residences and designing interior scenes. 
%
%
Recent years have seen significant advancements in the automated floorplan generation. Existing methods can be broadly categorized into rule-based and learning-based approaches. 
The former~\cite{merrell2010computer, liu2013constraint, laignel2021floor, shekhawat2021tool}, relying on specific user requirements and expert knowledge, typically optimize based on various explicit rules as constraints. This process is often sensitive to the modeling of constraints and the selection of parameters.
The latter~\cite{hu2020graph2plan, chaillou2020archigan, nauata2021house, para2021generative}, using deep neural networks, learns implicit design rules from real floorplans. While addressing the shortcomings of rule-based methods, it also introduces new issues:
(i) It is challenging to ensure that generated floorplans meet explicit constraints, such as pronounced misalignment; (ii) The generated results usually require heuristic post-processing to be converted into usable vector floorplans.

In this paper, we propose a novel framework called \textit{GSDiff} to directly synthesize vector floorplans. The core idea is to view vector floorplan synthesis as structural graph generation and decouple it into two tasks: wall junction generation and wall segment prediction. 
We represent the floorplan as a structural graph~\cite{sun2022wallplan}, where nodes represent wall junctions and edges represent wall segments. Additionally, to capture the floorplan semantics, we consider room labels as one of the node attributes. 
We first use a generative model based on a diffusion model to generate graph nodes, and then a predictive model based on a Transformer is used to determine graph edges between generated nodes, resulting in a complete structural graph.
To improve the design aesthetics, we also propose geometry-enhanced optimization techniques. During the node generation phase, we introduce a novel node alignment loss that optimizes the alignment error of nodes in mixed-base representations, which empowers the generative model to constrain node alignment. 
In the edge prediction phase, we employ an innovative edge perception enhancement strategy. This involves randomly interpolating a third point on the edges and self-supervising the model to predict the interpolation coefficients, which enhances the geometric perception ability of our edge prediction model, thereby improving the topological connectivity of structural graphs.
Finally, vector floorplans can be directly extracted from the generated structural graphs.

Extensive evaluations show that our method has significant advantages over state-of-the-art techniques on all metrics, enabling free generation and constrained generation. Our contributions are as follows: 
(i) A novel framework for automatically generating diverse, high-quality vector floorplans with various constraints by transforming the problem into a structured graph generation process.
(ii) An alignment error optimization strategy that improves node alignment for better node consistency in the node generation phase.
(iii) An innovative edge perception enhancement strategy that improves the edge accuracy in the edge prediction phase.

\begin{figure*}[t]
	\includegraphics[width=0.9\textwidth]{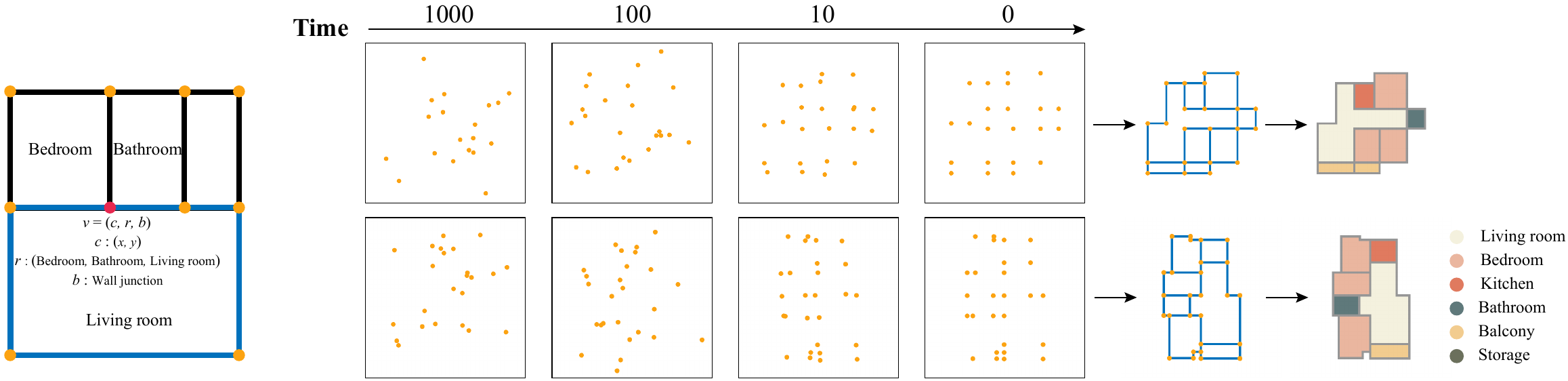}
	\put(-475, -8){\small (a) Floorplan representation}
	\put(-220, -8){\small (b) Generation process}
	\centering
	\caption{Overview. We represent the floorplan as a structural graph (a) and transform the vector floorplan synthesis into structural graph generation (b). We first generate graph nodes using a diffusion model, then predict the existence of edges between each pair of nodes, and finally extract rooms represented by polygons with semantic labels, resulting in vector floorplans.}
	\label{fig:teaser}
  \vspace{0mm}
\end{figure*}
\section{Related work}

Early works generate floorplans through rule-based floorplan optimization~\cite{merrell2010computer, liu2013constraint, wu2018miqp, laignel2021floor, wang2020generating, shekhawat2021tool, bisht2022transforming}. Due to the complexity of architectural design, deep learning methods have now become the mainstay of the field. Therefore, we focus on learning-based floorplan generation.

\paragraph{Imagery floorplan generation}

Some deep learning methods generate imagery floorplans. 
\textit{RPLAN}~\cite{wu2019data} proposes a two-stage method for floorplan generation that starts with predicting room locations and types, followed by detailing wall semantics and finalizing with vectorization to achieve the end floorplan. 
\textit{WallPlan}~\cite{sun2022wallplan} also converts floorplan generation as a graph generation task. However, it still generates floorplan images rather than vector formats via Convolutional Neural Networks (CNNs), therefore it can be categorized as an imagery generation.
These methods cannot operate as true generative models as they generate specific outputs from given inputs.
GAN-based floorplan generation~\cite{chaillou2020archigan, nauata2020house, nauata2021house} have gained traction for this purpose. However, these methods face challenges in generating structural elements, requiring complex post-processing to convert to vector formats.
In contrast, our method not only can generate multiple results from the same input but also bypasses such limitations by directly generating vector floorplans, simplifying the process and enhancing output usability.

\paragraph{Vector floorplan generation}

Vector floorplans are more widely used in practical applications. 
\textit{Graph2Plan}~\cite{hu2020graph2plan} introduces a method to create floorplans from specified bubble diagrams. However, aligning the boxes with the semantic representations requires complex post-processing to obtain vector floorplans.
~\cite{para2021generative} conceptualizes the floorplan as a box set, followed by optimization for geometric shaping. This method sometimes faces unsuitable constraints for optimization.
\textit{HouseDiffusion}~\cite{shabani2023housediffusion} represents an innovative application of diffusion models to floorplan generation, where floorplans are depicted as polygons with vertices categorizing rooms or doors. This method, however, encounters issues with room alignment, producing gaps or overlaps, and is limited by the necessity of specifying room categories and numbers up front.
Our work distinguishes itself by avoiding the limitations of box-set representations and directly producing vector floorplans. Furthermore, by utilizing a generative model, our method possesses the ability to generate diverse results from the same input, while Graph2Plan can only produce a single output.

\paragraph{Diffusion Model}

Diffusion models are a class of generative models to reverse the noise addition process, thereby enabling the generation of data that mimics the original distribution from Gaussian noise. A standard diffusion model generally contains the forward and backward processes~\cite{ho2020denoising}.
Diffusion models have made remarkable progress across various generation tasks, including image generation~\cite{nichol2021glide, ho2022cascaded, rombach2022high, saharia2022palette}, point cloud generation~\cite{nichol2022point}, and 3D model generation~\cite{poole2022dreamfusion, liu2023zero}. 
Our framework is centered on the diffusion model, and extends the capabilities of diffusion modeling for generating floorplans, enhancing the flexibility of the model to generate a wider variety of floorplans without the need for predefined room categories or quantities. 
\section{Method}

\begin{figure*}[t]
	\includegraphics[width=\linewidth]{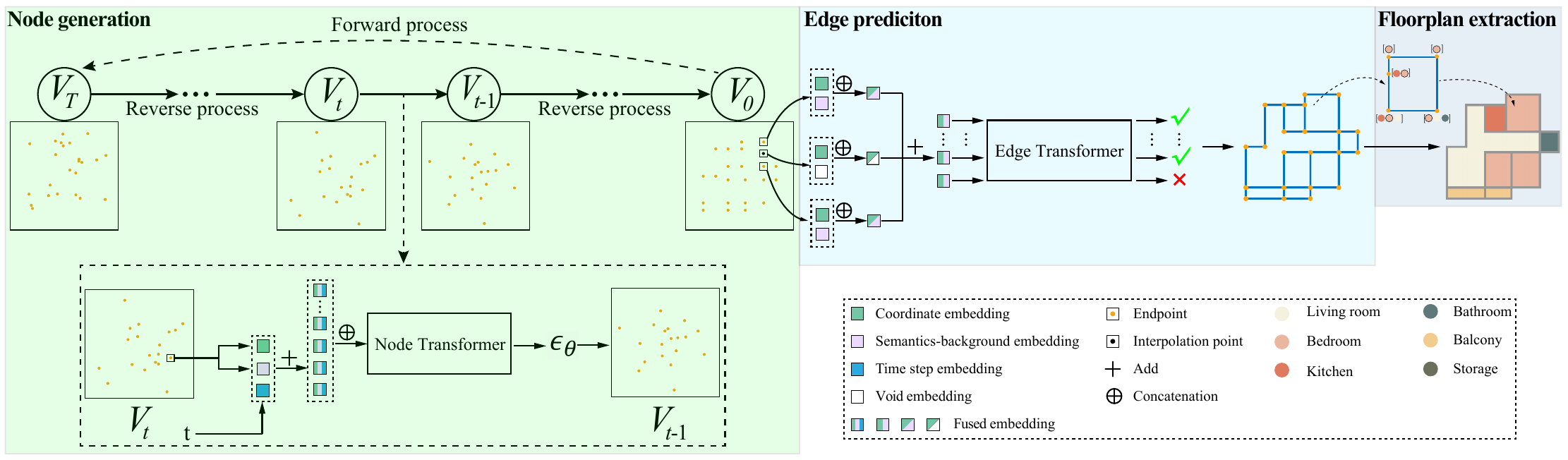}
	\centering
	\caption{Network architecture of~\textit{GSDiff}. we propose to decouple the structure graph generation into two stages: node generation and edge prediction, which results in the complete structural graph. Finally, all minimal polygonal loops of the structural graph are extracted as rooms to obtain the final vector floorplan.}
	\label{fig:DDPM}
  \vspace{0mm}
\end{figure*}

We convert floorplan generation into graph generation, by representing the vector floorplan as a structural graph \(G = (V, E)\). Directly generating structural graphs \(P(G)\) is pretty complex. To this end, we propose a novel generation framework~\textit{GSDiff} (Figure~\ref{fig:teaser}) to decouple the structure graph generation into two stages: node generation and edge prediction, i.e., \(P(G) = P(V, E) = P(V)P(E|V)\), which results in the complete structural graph. Finally, all minimal polygonal loops of the structural graph are extracted as rooms to obtain the final vector floorplan. 
~\textit{GSDiff} takes design constraints, such as floorplan boundaries, as input and produces high-quality vector floorplans. For simplicity, we first introduce unconstrained generation and more details about constrained generation will be presented later. 

\subsection{Floorplan representation}

We represent the vector floorplan as a structural graph (Figure~\ref{fig:teaser} (a)): wall junctions are abstracted as nodes, and wall segments as edges. \(G = (V, E)\), where \(V = \{v_1, \ldots, v_n\}\) is the node set, and \(E = \{(v_i, v_j) | v_i, v_j \in V\}\) is the edge set.
\(v_i = (c_i, r_i, b_i)\), where \(c_i = (x_i, y_i) \in \mathbb{R}^2\) is the positional coordinate in the range of $[-1, 1)$. \(r_i \in [0, 1]^n\) denotes the semantics of all rooms that surround \(v_i\), where \(0\) for absence and \(1\) for presence of the specific room category, and \(n\) is the number of room categories, which is \(7\) (\textit{Living room}, \textit{Bedroom}, \textit{Kitchen}, \textit{Bathroom}, \textit{Balcony}, \textit{Storage}, \textit{External area}) in our experiments. 
Let \(F = \{R_1, R_2, \ldots, R_m\}\) represent all polygonal loops of rooms, each loop \(R_i = \{v_{i,1}, v_{i,2}, \ldots, v_{i,N_i} | v_{i,j} \in V\}\) is defined by a sequence of graph nodes. The room category of \(R_i\) is the shared room category of all graph nodes forming that room.
Different from \cite{sun2022wallplan}, we set a fixed node number
\SZ{\(N=53\)}
by considering background nodes\SZ{, which is determined by the maximum number of nodes in floorplans in our dataset.} Background nodes are filtered out by assigning a background attribute for each node, denoted as \(b_i \in \{0, 1\}\), where 0 denotes the junction node and 1 for the background node.
Vector floorplans have a multi-level structure. Spatially, 0D wall junctions form 1D wall segments, which close into 2D rooms. Semantically, each wall junction that forms a room has the same room semantics. Therefore, generating all 0D wall junctions basically determines the whole floorplan structure. 

\subsection{Node Generation}

We adopt a diffusion model~\cite{ho2020denoising} based network architecture to generate nodes (Figure~\ref{fig:DDPM}). 

\paragraph{Network architecture} 

Our forward process incrementally adds noise to an original data sample $V_0$ over $T$ steps, resulting in a sample resembling Gaussian noise. 
Our reverse process is a Transformer~\cite{vasvani2017attention} based neural network, which takes the noisy node set at time $t$ and outputs the predicted noise $\epsilon_{\theta}$ at time $t-1$ ($\theta$ represents the parameters of Transformer), therefore inferring the noisy node set at time $t - 1$. When the time step reaches 0, the node generation process terminates.
Given a noisy node set $V_t$ at time $t$. We initialize the node embedding of node $f_i$ as
\begin{equation}
\label{node-fi}
f_i = f^{\text{c}}_i + f^{(r, b)}_i + f^t_i
\end{equation}
$f^{\text{c}}_i = [\gamma(x_i), \gamma(y_i)] \in \mathbb{R}^d$ is the coordinate embedding, where $\gamma(\cdot)$ is the positional encoding~\cite{vasvani2017attention}. $f^{(r, b)}_i \in \mathbb{R}^d$ is the embedding of $(r_i, b_i)$ using a fully connected layer. $f^t_i \in \mathbb{R}^d$ is the time embedding using a feed-forward neural network (FFN). 

\paragraph{Loss function}

The reconstruction loss is typically defined as the Mean Squared Error (MSE) for training:
\begin{equation}
MSE(\epsilon, \epsilon_{\theta}) = \mathbb{E} \left[ \| \epsilon - \epsilon_{\theta}(V_t, t) \|^2_2 \right]
\end{equation}

Due to the probabilistic nature of neural networks, the generated nodes may not be perfectly aligned.  We propose a new alignment loss that optimizes alignment errors.
Unlike natural language or images, structural graphs have precise geometric relationships, like perfectly horizontal or vertical walls. Directly regressing real-valued coordinates often fails to capture this precision. Thus, we convert real values into binary representations for regression, which discretizes the continuous coordinate space for more precise learning. However, learning discrete representations accurately is challenging, as errors in higher-order bits can cause significant real-value errors.
To address this, we propose mixing multiple radix representations. Real-valued representations can heavily penalize large errors but are lenient on small misalignments. Binary representations, though overly discrete and causing significant penalties for small misalignments, are ineffective for large errors. By ``interpolating" various radix representations, we aim for a smooth transition that penalizes large errors appropriately while remaining sensitive to small misalignments, balancing both advantages for better performance.
We aim to enhance node alignment by applying the above concepts to propose a novel alignment loss across multiple bases, including real, binary, quaternary, octal, and hexadecimal:
\begin{equation}
MixAlg(\hat{V}_0) = Alg(\hat{V}_0) + \sum_{k\in{2,4,8,16}} Alg^{k}(\hat{V}_0)
\end{equation}
%
\begin{equation}
\label{8}
\text{Alg}(\hat{V}_0) = \sum_{i=1}^n g(\min \left( \Delta b_i^X, \Delta b_i^Y \right))
\end{equation}
%
%
\begin{equation}
\label{9}
Alg^{2}(\hat{V}_0) = \sum_{i=1}^{n} g^{2}(\sum_{j=1}^{s} \left(Base^{2} \left( \min \left( \Delta b^X_i, \Delta b^Y_i \right) \right) \right)_j)
\end{equation}
where \(Base^{k}(\cdot)\) is the $k$-base representation, \( \Delta b^*_i = \min_{j \neq i} | b^*_i - b^*_j | \), \( * \in \{X, Y\} \), \( g(x) = -2* \log \left( 1 - \frac{x}{2} \right) \), \( g^{k}(x) = -d^{k} \log \left( 1 - \frac{x}{d^{k}} \right) \), with \( d^k \) indicating the maximum allowable distance under $k$-base. \(n\) is the node number, \(s\) is the bit size. With $k = 2, 4, 8, 16$, $s =  12, 6, 5, 3$ and $d^k = 12, 18, 35, 45$. \( (\cdot)_i \) denotes the \( i \)-th bit.

We combine the reconstruction loss and alignment loss:
\begin{equation}
\mathcal{L} = MSE(\epsilon, \epsilon_{\theta}) + \omega(t) MixAlg(\hat{V}_0)
\end{equation}
We adopt time-related weighting scheme \(\omega(t)\)~\cite{chen2024towards}, assigning higher weights at smaller time step \(t\). More details can be found in the supplementary material.

\subsection{Edge Prediction}

We use a Transformer-based predictive model to determine graph edges between generated nodes (Figure~\ref{fig:DDPM}).

\paragraph{Network architecture} 

For each candidate edge \((v_i, v_j)\), the input embedding is obtained by fusing the embeddings of \(v_i\) and \(v_j\): 
\begin{equation}
f_{i, j} = f_i + f_j 
\end{equation}
The coordinate embedding \(f^c_i\in \mathbb{R}^{\frac{d}{2}}\) and the semantic-background embedding \( f^{(r,b)}_i \in \mathbb{R}^{\frac{d}{2}}\) are concatenated as the node embedding \(f_i = f^c_i\oplus f^{(r,b)}_i\).

To enhance the model's robustness, we introduce noise to the node features during the training phase. Specifically, for the normalized 2D coordinate of each node, we add truncated Gaussian noise \(\epsilon^{\prime} \sim Truncate\left(\mathcal{N}(0, \sigma_c^2), -3\sigma_c, 3\sigma_c\right)\), which sampled from Gaussian noise but is bounded at both ends. For the semantic attributes of the nodes, we randomly flip each bit in the multi-hot representation with a probability \( p_{flip} \), simulating label noise. We set \( \sigma_c = 1 \) and \( p_{flip} = 0.01 \).

\paragraph{Loss function}

Edge prediction is essentially geometric inference, rich geometric information will be more helpful. However, the geometric information of an edge includes more than just its endpoints. To improve edge perception, we propose an edge perception enhancement strategy. Specifically, we add a random interpolation point and require the model to predict its interpolation coefficient, enhancing the model's ability to infer intermediate edge structures.
For each edge \((v_i, v_j)\), the interpolation point's coordinates and attributes are defined as:
\begin{equation}
c_\lambda = \lambda c_i + (1 - \lambda)c_j \in [-1, 1)^2
\end{equation}
where $r_\lambda = \textbf{0} \in \mathbb{R}^7$, \( \lambda \sim U(0, 1) \) is a random interpolation coefficient. The supervised loss for interpolation is:
\begin{equation}
\mathcal{L}_{\lambda} = \mathbb{E} \left[ \left| \tilde{\lambda} - \hat{\lambda}_\theta(e_{ij}) \right| \right]
\end{equation}
where \( \hat{\lambda}_\theta \) represents the model's predicted interpolation coefficient. $\tilde{\lambda}=1-\lambda$, if $\lambda>0.5$; $\tilde{\lambda}=\lambda$, if $\lambda \leq 0.5$. 

The enhanced edge feature includes the features of the two endpoints of an edge, as well as the random interpolation point. The final edge prediction loss for training is:
\begin{equation}
\mathcal{L}_{edge} = \mathcal{L}_{cls} + \mathcal{L}_{\lambda}
\end{equation}
where \(\mathcal{L}_{cls} \) is the Cross-entropy classification loss. More details are provided in the supplementary material.

\paragraph{Floorplan extraction}

So far, we have obtained the structural graph \(G = (V, E)\). We can simply extract all minimal polygonal loops as rooms. Considering that the prediction based on neural networks might contain errors, leading to the absence of a category shared by all nodes, we select the most frequently occurring category as the room type. 
If multiple categories have the highest frequency, we consider factors such as the rarity of the category and determine the room type based on the following priority: \textit{Storage} $>$ \textit{Bathroom} $>$ \textit{Kitchen} $>$ \textit{Bedroom} $>$ \textit{Balcony} $>$ \textit{Living room}. An illustration is provided in the supplementary material.

\begin{figure}[t]
	\includegraphics[width=\linewidth]{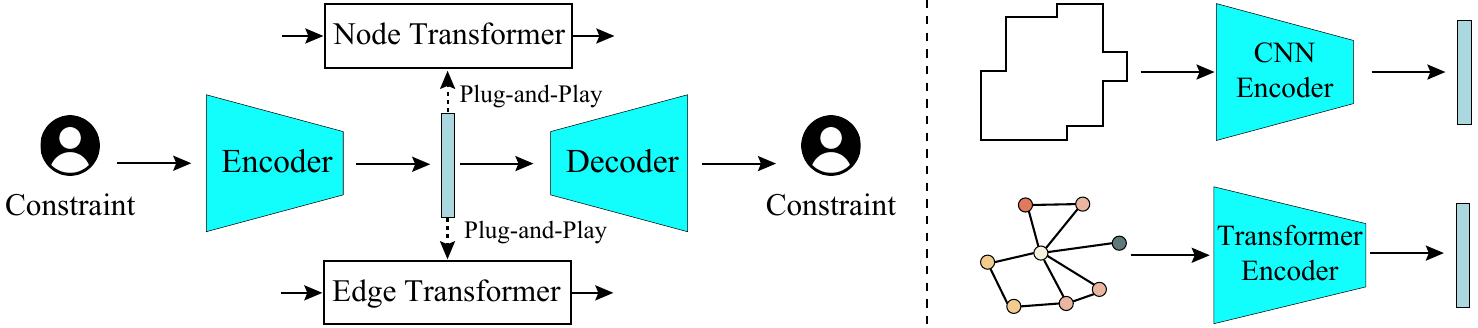}
	\centering
	\caption{Constrained generation. Embeddings for constraints are obtained through respective encoders and used as inputs for node generation and edge prediction. Topology constraints use a Transformer-based encoder, while boundary constraints use a CNN-based encoder.}
	\label{fig:constrained}
\end{figure}

\subsection{Constrained Generation}

The proposed framework supports constrained floorplan generation. To incorporate constraints into our framework, we introduce constraint encoders to guide the generation. We encode the constraints using an encoder specific to the constraint modality, and the encoded features serve as input to the decoder to model the conditions via cross-attention.
To improve the encoding capability of constraint encoders, we train each encoder individually. Specifically, we use a ``pre-training + fine-tuning" paradigm, where we first pre-train the constraint encoder on the synthetic constraint data, and then fine-tune the encoder on the real constraint data of the dataset to achieve better generalization. We train constraint encoders in the framework of autoencoder. Without loss of generality, we focus on the boundary-constrained generation and topology-constrained generation in this paper (Figure~\ref{fig:constrained}). See more details in the supplementary material. 

\paragraph{Boundary-constrained generation}

A boundary refers to the outer contour of a floorplan, typically represented as a polygon. To encode a boundary, we draw the boundary polygon on an image with a resolution of $256\times256$, converting the boundary polygon into an image. We then use a CNN-based encoder for encoding. Specifically, we modify U-Net~\cite{unet} by removing skip connections and adding residual connections, as our boundary encoder. The encoder outputs a feature map of $16\times16$ with $1024$ channels. During the pre-training phase, we generate random polygons on the image and learn their encodings with a CNN-based autoencoder. In the fine-tuning phase, we train with real boundary data from the dataset.
The boundary embeddings are fed to node and edge Transformers, ensuring that the boundary of the generated structural graph adheres to the given constraint of the boundary.

\begin{figure*}[t]
	\centering
	\includegraphics[width=0.9\textwidth]{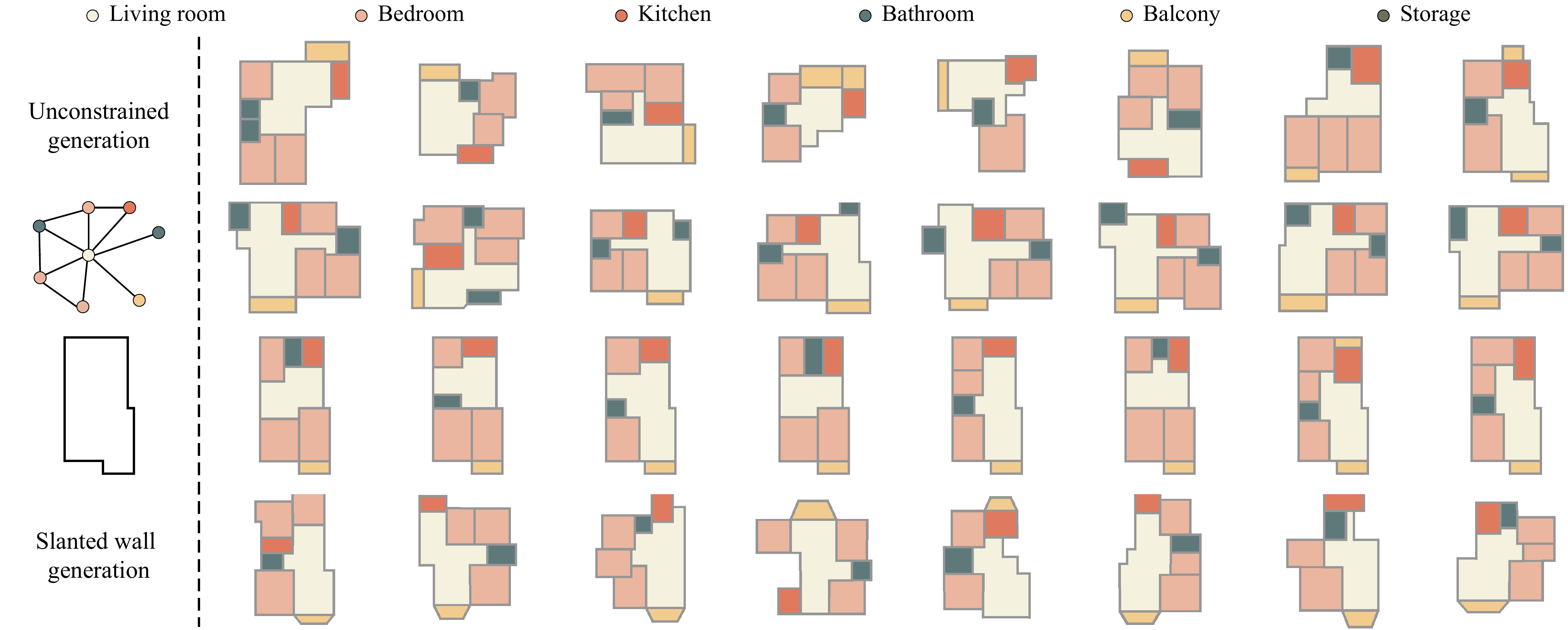}
	\caption{A gallery of vector floorplans generated using our framework. From top to bottom: unconstrained generation, topology-constrained generation, boundary-constrained generation, and unconstrained generation with slanted walls.}
	\label{fig:results}
\end{figure*}

\paragraph{Topology-constrained generation}

Floorplan topology is used to describe the connectivity between rooms. It is an undirected graph where each node represents a room, and each edge represents a connectivity between two rooms. We encode the topological graph with a Transformer, which outputs 256D embeddings of all rooms, constituting topological embeddings. During the pre-training phase, we randomly generate topological graphs and learn their embeddings using a Transformer-based autoencoder. During the fine-tuning phase, we train with real topological graph data. 
The topology embeddings are fed to node and edge Transformers, ensuring that the topology of the generated structural graph adheres to the given constraint of the topology. 

\section{Experiments}

Our method is implemented using Pytorch and trained on an NVIDIA GeForce GTX 4090 GPU. To ensure the quality of training at each stage, we train each network separately, using the Adam optimizer~\cite{kingma2014adam} with an initial learning rate of \(1 \times 10^{-4}\). 
We have used the \textit{RPLAN} dataset~\cite{wu2019data} for training and testing, which contains more than 80K residential floorplans with dense annotation. The sample size for validing and testing is 3,000 each and the rest is used for training.
Creating a vector floorplan takes an average of 0.17 seconds without constraints, 0.67 seconds with boundary constraints, and 0.86 seconds with topological constraints.
See more in the supplementary material. 

\subsection{Qualitative Evaluation}

\paragraph{Unconstrained generation} 

Unconstrained generation means that diverse floorplans can be generated without any inputs. The unconstrained generation allows users to explore freely, potentially inspiring more creative and innovative designs. It is worth noting that less research work is currently focused on unconstrained floorplan generation. By tilting balcony walls in the dataset, our method can also generate floorplans with slanted walls. Thanks to our robust structural graph representation, alignment error optimization strategy, and edge perception enhancement strategy, we can generate diverse, high-quality, vector floorplans without any inputs (Figure \ref{fig:results}). For more results, please refer to the supplementary materials.

\begin{figure}[t]
	\centering
	\includegraphics[width=0.48\textwidth]{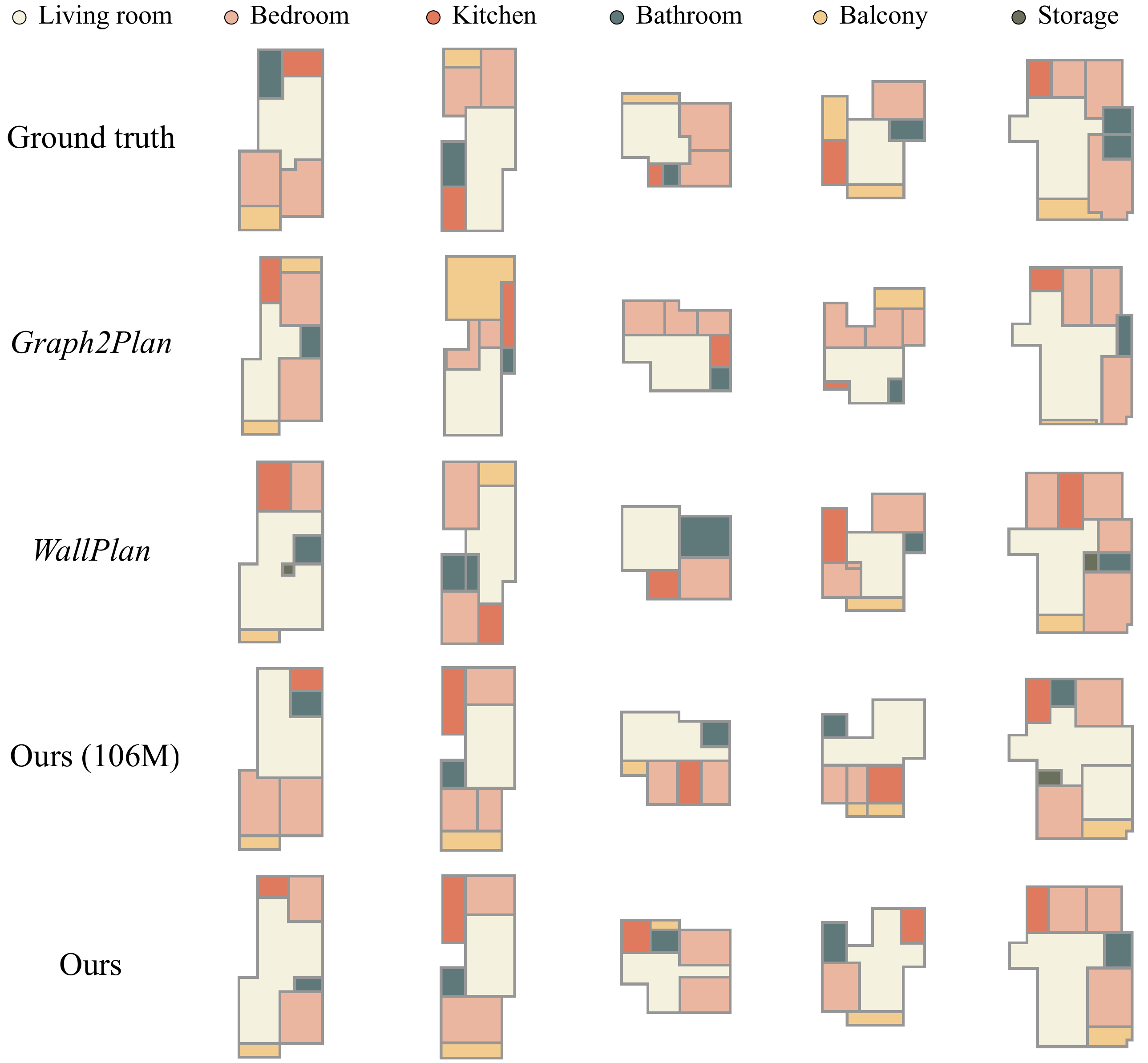}
	\caption{Comparison with the ground truth, \textit{Graph2Plan}, and \textit{WallPlan} on the boundary-constrained generation. Our method can produce more reasonable floorplans.}
	\label{fig:comparison-boundary}
\end{figure}

\paragraph{Boundary-constrained generation} 

Both \textit{Graph2Plan} and \textit{WallPlan} can generate floorplans from boundaries, thus we compare our method against them. Figure \ref{fig:comparison-boundary} shows a comparison of floorplans generated by different methods. 
\textit{Graph2Plan} is prone to generating issues such as unreasonable space divisions and areas, making these defects particularly noticeable as almost every sample exhibits significant flaws. Specifically, the second column features a huge balcony and tiny bedrooms, kitchen, and bathroom; the third column lacks a balcony; the fourth column has a bedroom accessible only through another bedroom on the right, and the fifth column includes an overly narrow balcony.
\textit{WallPlan}, although it produces fewer unreasonable shapes than \textit{Graph2Plan}, some defects still persist. In the first column, the storage cabinet next to the bathroom should be against a wall, but it is located in the middle of the room; the second column has an unreasonable bathroom division, the third column has overly simplistic divisions, a huge bathroom, and a missing balcony; the fourth column has a super small, impractical bedroom, and the fifth column has a bathroom blocked by storage.
The limitations of \textit{Graph2Plan} and \textit{WallPlan} are that they can only generate a single result for a specific input, and CNNs struggle to model long-range semantic relationships that involve the reasonableness of room layouts. In contrast, our model, benefiting from structural representation and attention mechanisms, can produce a variety of results that are closer to the fundamental facts of actual buildings.

\begin{figure}[t]
	\centering
	\includegraphics[width=0.48\textwidth]{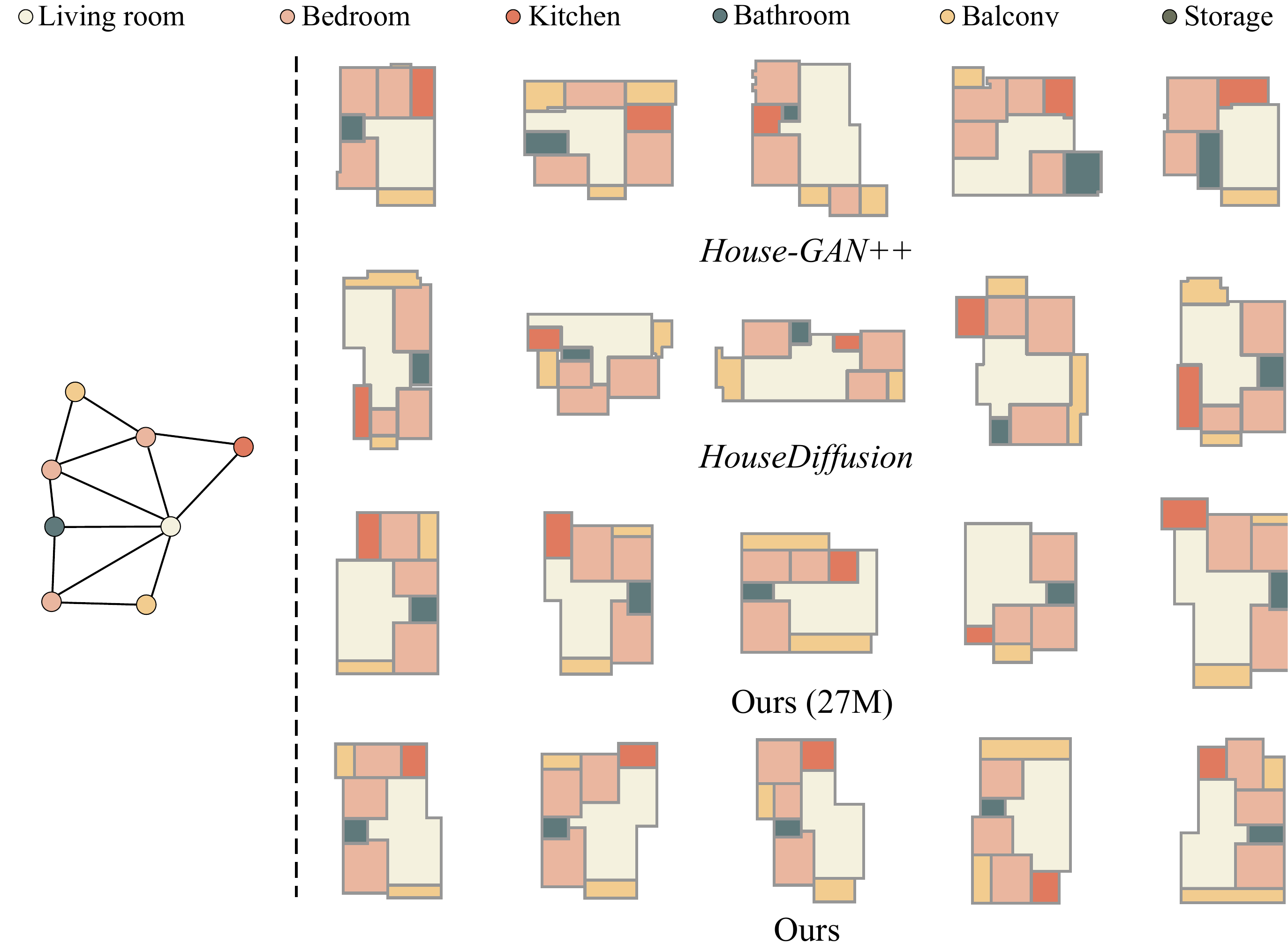}
	\caption{Comparison with \textit{House-GAN++} and \textit{HouseDiffusion} on the topology-constrained generation. Our results exhibit greater diversity, better adherence to constraints, and superior visual quality.}
	\label{fig:comparison-topo-v1}
\end{figure}

\paragraph{Topology-constrained generation} 

We compare our method with \textit{House-GAN++} and \textit{HouseDiffusion}, which can generate floorplans based on topology graph constraints. It's noteworthy that \textit{HouseDiffusion} requires the number of vertices for each room polygon to be pre-specified, hence we retrieve samples from the dataset as its input. Figure~\ref{fig:comparison-topo-v1} shows a comparison of different methods.
The principal issue with \textit{House-GAN++} lies in the peculiar room shapes, and jagged room boundaries are prevalent, leading to weaker visual aesthetics. Specifically, the overly small balcony at the top of the first column, the unreasonable arrangement between balconies and bedrooms in the second, third, and fourth columns, and the bedroom obstructed by the bathroom in the fifth column. And nearly every sample fails to meet the constraints.
\textit{HouseDiffusion} generates better quality compared to \textit{House-GAN++}, yet it requires the number of vertices for each room polygon to be pre-specified as an extra ``constraint", limiting the diversity of room shapes. Moreover, limited by their binary coordinate optimization, there are issues with boundaries not aligning well, preventing the acquisition of a good wall structure. Many of their generated results fail to satisfy the room's topological constraints. Specifically, the poorly shaped balconies in the first, third, and fourth columns, the bathroom in the middle of the house in the second column, and the kitchen in the fourth column that can only be entered from a bedroom. 
In contrast, our method can generate high-quality and diverse room boundaries, making the generated results more natural, aesthetically pleasing, and compliant with constraints.

\subsection{Quantitative Evaluation}

\begin{table*}[t]
	\begin{center}
		\caption{Quantitative evaluation. \SZ{\# Param: parameter counts.} Each experiment is repeated 5 times to eliminate randomness, and the average results are reported. The smaller the better for all metrics. \nothing{The best results are highlighted in \textbf{bold}.}\SZ{The color \textcolor{cyanTop}{cyan}
and \textcolor{magentaTopTwo}{magenda} mark the top-two results.} Void cells: GED is only applicable to topological graph constraints.}
		\centering
		\label{tab:table1}
		\renewcommand\arraystretch{1.1}
		\setlength{\tabcolsep}{2mm}
		\scalebox{0.93}
		{
			\begin{tabular}{c|cc|ccc|cccccc}
				\toprule[2pt]
				Constraint & Method & \# Param & FID & KID & GED & Living & Kitchen & Bedroom & Storage & Bathroom
				& Balcony\\
				\midrule
				\multirow{4}*{\hfil Topology} & \textit{House-GAN++} & 2M & 48.40 & 54.66 & 3.9 & 0.860 & 1.300 & 0.984 & \textcolor{cyanTop}{1.151} & 1.378 & 1.037\\
				& \textit{HouseDiffusion} & 27M & 11.87 & 7.23 & 2.59 & 0.955 & 0.971 & 0.979 & 0.567 & \textcolor{cyanTop}{0.967} & \textcolor{cyanTop}{0.994}\\
				& Ours (27M) & 27M & \textcolor{cyanTop}{6.64} & \textcolor{cyanTop}{1.62} & \textcolor{magentaTopTwo}{0.57} & \textcolor{magentaTopTwo}{0.977} & \textcolor{magentaTopTwo}{0.981} & \textcolor{magentaTopTwo}{0.989} & \textcolor{magentaTopTwo}{0.817} & 0.954 & {0.966}\\
                & Ours & 125M & \textcolor{magentaTopTwo}{6.82} & \textcolor{magentaTopTwo}{1.79} & \textcolor{cyanTop}{0.49} & \textcolor{cyanTop}{0.992} & \textcolor{cyanTop}{0.991} & \textcolor{cyanTop}{1.004} & 0.752 & \textcolor{magentaTopTwo}{0.966} & \textcolor{magentaTopTwo}{0.976}\\
				\midrule
				\multirow{4}*{\hfil Boundary} & \textit{Graph2Plan} & 8M & 8.40 & 1.34 & - & 1.034 & \textcolor{magentaTopTwo}{0.975} & 1.013 & \textcolor{magentaTopTwo}{0.727} & 0.959 & 0.859 \\
				& \textit{WallPlan} & 106M & 9.07 & 1.02& - & 0.992 & 1.033 & 0.923 & \textcolor{cyanTop}{1.136} & 1.058 & \textcolor{cyanTop}{1.008} \\
                & Ours (106M) & 106M & \textcolor{magentaTopTwo}{7.83} & \textcolor{cyanTop}{0.51} & - & \textcolor{cyanTop}{1.000} & {0.954} & \textcolor{cyanTop}{1.007} & {0.500} & \textcolor{magentaTopTwo}{0.964} & \textcolor{magentaTopTwo}{0.977}\\
                & Ours & 137M & \textcolor{cyanTop}{7.50} & \textcolor{magentaTopTwo}{0.56} & - & \textcolor{magentaTopTwo}{1.007} & \textcolor{cyanTop}{0.990} & \textcolor{magentaTopTwo}{0.992} & 0.454 & \textcolor{cyanTop}{0.967} & 0.935\\
				\toprule[2pt]
			\end{tabular}
		}
	\end{center}
\end{table*}

\paragraph{Distribution comparison}

The distribution comparison is used to analyze the overall generation capability of a generative model by comparing the differences between the distributions of generated data and real data. We consider the following representative metrics:
(1) FID (\emph{Fr'echet Inception Distance})\cite{heusel2017gans} is used to measure the quality and diversity of generated images by calculating the distribution distance between real and generated data in feature space; (2) KID (\emph{Kernel Inception Distance})\cite{binkowski2018demystifying} uses kernel methods to calculate the maximum mean discrepancy in feature space and is generally considered to be more robust than FID.
Table~\ref{tab:table1} shows the results of the distribution comparison. It's worth mentioning that we found that the image metrics FID and KID are influenced by a combination of factors such as room type, area, room layout, wall shape, and alignment, which can measure the generation results on the whole, making them excellent indicators for floorplan generation.
In the evaluation of boundary-constrained generation, we selected the intersection of the test sets of our model, \textit{Graph2Plan}, and \textit{WallPlan} (378 boundaries in total) as input constraints, generating one sample per boundary for each method (as \textit{Graph2Plan} and \textit{WallPlan} can only produce a single output). Table~\ref{tab:table1} shows that the visual, geometric, and other features of our generation results outperform \textit{Graph2Plan} and \textit{WallPlan}.
For the evaluation of topology graph-constrained generation, we selected the intersection of the test sets of our model, \textit{House-GAN++}, and \textit{HouseDiffusion} (a total of 757 topology graphs) as input constraints, generating 757 * 5 samples for each method, calculating the results five times and taking the average. Table~\ref{tab:table1} indicates that our generation results maintain better geometric consistency, visual appeal, and better practicality.
Additionally, for topology-constrained generation, we reference \textit{HouseDiffusion}~\cite{shabani2023housediffusion}, and introduce Graph Edit Distance (GED)~\cite{abu2015exact} as an additional metric for evaluation. GED is a graph-matching approach that calculates the distance between the input bubble diagram and the one reconstructed from the generated floorplan. For GED of \textit{House-GAN++}, we directly use the reported GED in their paper. Table~\ref{tab:table1} indicates that our generation results maintain better constraint satisfaction, which benefits from our structural representation. 
For the unconstrained generation of floorplans with slanted walls, FID=12.02, KID=9.98.

\paragraph{Statistics comparison}

We also conducted a statistical analysis of the generated vector results to evaluate the quality of the generated floorplans in terms of practicality. We ran the test set five times for each method. For each method's generated results, we calculated the amount of each type of room. We calculated the average values of these statistics and compared them with the corresponding statistics of the ground truth in the real dataset (the closer to 1, the better). The results, as shown in Table~\ref{tab:table1}, indicate that our method has a clear advantage in practicality compared to state-of-the-art techniques. For most types of rooms (living room, kitchen, bedroom), the amount generated by our method is closer to the real dataset. Only the balcony, storage, and bathroom are closer to the dataset by \textit{House-GAN++} and \textit{HouseDiffusion}, but the difference in closeness with ours is not significant. Moreover, for boundary constraints, our method also shows an advantage in bathrooms.

\subsection{Ablation Study}

\begin{table}[t]
	\begin{center}
		\caption{Ablation Study. FE (Fake Edge) is the ratio of misclassified edges, and AE (Alignment Error) is the node alignment error. Each experiment is repeated 5 times to eliminate randomness, and the average results are reported. The smaller the better for all metrics. The best results are highlighted in \textbf{bold}. }
		\centering
		\label{tab:quantitative_evaluation}
		\renewcommand\arraystretch{1.1}
		\setlength{\tabcolsep}{2mm}
		\scalebox{1}
		{
			\begin{tabular}{ccccc}
				\toprule[2pt]
				Method & FID & KID & FE(\%) & AE \\
				\midrule
				\textbf{Node\(_{None}\) + Edge\(_{None}\)} & 6.23 & 4.12 & 1.35 & 0.34 \\
				\textbf{Node\(_{Real}\) + Edge\(_{None}\)} & 6.18 & 4.16 & 1.32 & 0.29 \\
				\textbf{Node\(_{Binary}\) + Edge\(_{None}\)} & 5.90 & 4.03 & 1.28 & 0.33 \\
				\textbf{Node + Edge\(_{None}\)} & \textbf{4.96} & \textbf{2.92} & \textbf{1.15} & \textbf{0.23} \\
				\midrule
				\textbf{Node\(_{None}\) + Edge} & 5.72 & 3.74 & 1.13 & 0.33 \\
				\textbf{Node\(_{Real}\) + Edge} & 5.84 & 3.93 & 1.11 & 0.29 \\
				\textbf{Node\(_{Binary}\) + Edge} & 5.84 & 3.96 & 1.13 & 0.34 \\
				\textbf{Node + Edge} (Ours) & \textbf{4.83} & \textbf{2.84} & \textbf{0.95} & \textbf{0.23} \\
				\toprule[2pt]
			\end{tabular}
		}
	\end{center}
\end{table}

We introduce two geometric enhancement strategies: one for alignment enhancement that optimizes the alignment error of nodes in mixed-base representations, and the other for perception enhancement that enhances the geometric perception ability of our edge prediction. 
To evaluate these two strategies, we have conducted a series of ablation experiments (Table~\ref{tab:quantitative_evaluation}). In these experiments, \textbf{Node\(_{None}\)} indicates no alignment enhancement for node generation, \textbf{Node\(_{Real}\)} indicates the only use of continuous real-valued alignment error optimization, \textbf{Node\(_{Binary}\)} indicates the only use of binary discrete alignment error optimization, \textbf{Edge\(_{None}\)} indicates no perception enhancement for edge prediction, and \textbf{Node} and \textbf{Edge} is our full methods.

Our method (\textbf{Node + Edge}) outperforms the other ablation methods in terms of all evaluated metrics, demonstrating that our geometric enhancements effectively improve the quality of the generated floorplans.
For FID and KID, our method achieved the lowest FID=4.83 and KID=2.84, indicating that the generated floorplans are more similar to the real data distribution. This demonstrates that both geometric alignment and edge perception enhancements contribute to higher visual fidelity and distribution alignment. FE is minimized with our method (0.95\%), which means the structural integrity and accuracy of the generated floorplans are effectively improved. AE remains consistently low (0.23), highlighting that our method effectively maintains geometric consistency and precision in the generated floorplans.

\SZ{
\subsection{Retrieval Analysis}

To further evaluate the generalization ability of the model, we performed a comprehensive retrieval analysis of the generated floorplans against those in the training set. For each generated floorplan (3,000 in total), we compute the minimum Euclidean distance and Wasserstein distance in image space between it and all floorplans in the training set: if the training set sample corresponding to the minimum distance is visually similar to the generated floorplan,  it indicates possible overfitting. The retrieval results show that there is a significant difference between the results we generate and those retrieved from the dataset, i.e., our model has good generalization performance. 
However, the distance values are not intuitive enough. 
To illustrate the difference intuitively, we provide several examples in Fig. ~\ref{fig:retrieval_examples} showing the closest match between the generated floorplans and the ones in the training set.

\begin{figure}[t]
	\centering
	\includegraphics[width=0.48\textwidth]{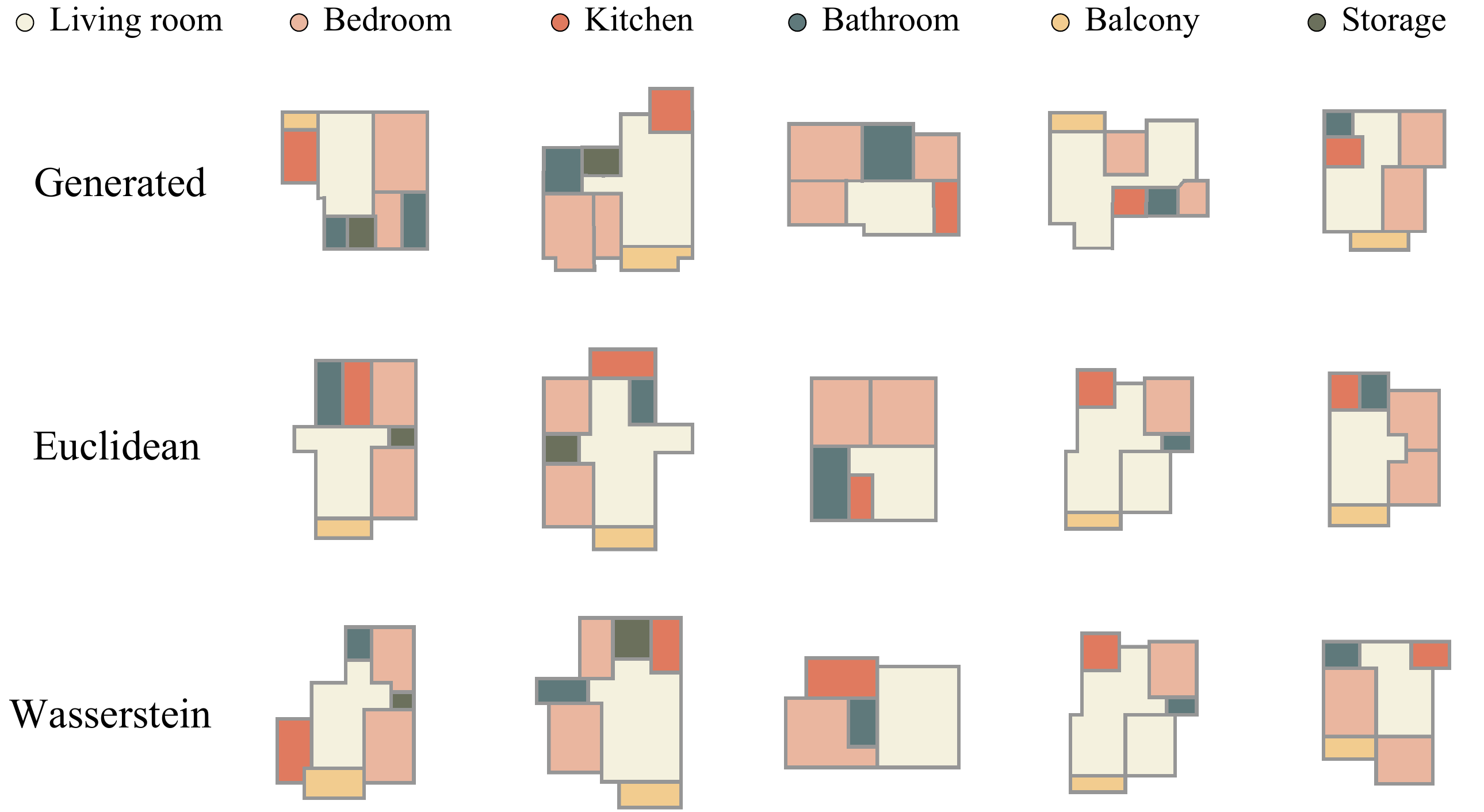}
	\caption{Retrieval Analysis. Generated: the generated floorplans. Euclidean(Wasserstein): the closest training floorplans based on Euclidean(Wasserstein) distance. 
    }
	\label{fig:retrieval_examples}
\end{figure}

\subsection{Discussion on Model Sizes} 

To ensure a fair comparison, we evaluated our method under equivalent model parameter sizes.

\paragraph{Topology-constrained generation}

Our original model for topology-constrained generation has approximately 125 million parameters, while \textit{House-GAN++} and \textit{HouseDiffusion} have about 2 million and 27 million parameters, respectively. To assess the impact of model size, we reduced our model's parameter count to 27 million to match that of \textit{HouseDiffusion}, retraining it with all other configurations unchanged. As shown in Figure~\ref{fig:comparison-topo-v1} and Table~\ref{tab:table1}, even with the same parameter size, our method still achieves superior performance. This indicates that the effectiveness of our method stems from its inherent advantages rather than a larger parameter count. Regarding \textit{House-GAN++}, we could not adjust our model to match their 2 million parameters. However, \textit{House-GAN++} has inherent problems. See the supplementary material for details.

\paragraph{Boundary Constrained Generation}

Our model has about 137 million parameters. In comparison, \textit{WallPlan} has about 106 million parameters and \textit{Graph2Plan} has about 8 million parameters. To evaluate the impact of model size, we reduce the number of parameters of our model to 106 million to match \textit{WallPlan} and retrain it. As shown in Fig.~\ref{fig:comparison-boundary} and Table~\ref{tab:table1}, our method still achieves superior performance, which also shows that our method has its inherent advantages. Regarding \textit{Graph2Plan},  we were unable to adjust the model to match their 8 million parameters. However, \textit{Graph2Plan} has inherent problems. See the supplementary material for details.

\subsection{Evaluation on Other Datasets}

To further demonstrate the generalization and effectiveness of our model, we conducted experiments on the \textit{LIFULL} dataset~\cite{lifull}. We obtained the vectorized subset of \textit{LIFULL} dataset via \textit{Raster-to-Graph}~\cite{raster-to-graph}, which contains 10,804 floorplans: 500 are used for validation, 500 for testing, and the rest for training. We trained our unconstrained model using the same configurations as with \textit{RPLAN}, adjusting only the room categories to align with the 12 in \textit{LIFULL}. Table~\ref{tab:other-evaluation} and Fig.~\ref{fig:other-dataset} showcases the quantitative and qualitative results, respectively, demonstrating our method's generalizability and effectiveness in diverse scenarios. However, as the \textit{Raster-to-Graph} authors themselves mentioned, the annotation quality of the \textit{LIFULL} vectorized dataset they extracted is weaker and the number is much smaller than that of \textit{RPLAN}, which make the effect on \textit{LIFULL} limited. Possible future works would include extending our method to the \textit{MSD}~\cite{vanengelenburg2024msd} dataset, which contains a significant share of layouts of multi-apartment dwellings. \textit{MSD} lacks vectorized structural graphs, and the presence of complex shapes of walls in the images makes vectorized structural graphs difficult to extract.

\begin{table}[t]
    \caption{Quantitative evaluation on the \textit{LIFULL} dataset.}
    \centering
    \label{tab:other-evaluation}
    \renewcommand\arraystretch{1.1}
    \setlength{\tabcolsep}{2mm}
    \newcolumntype{C}{>{\centering\arraybackslash}X}
    \newcolumntype{P}[1]{>{\centering\arraybackslash}p{#1}}
    \begin{tabularx}{\columnwidth}{P{0.25\columnwidth}CCCC}
        \toprule[2pt]
        Dataset & FID & KID & FE(\%) & AE \\
        \midrule
        {\textit{LIFULL}} & {12.44} & {3.61} & {6.11} & {3.75} \\
        \bottomrule[2pt]
    \end{tabularx}
\end{table}

\begin{figure}[t]
	\centering
	\includegraphics[width=0.48\textwidth]{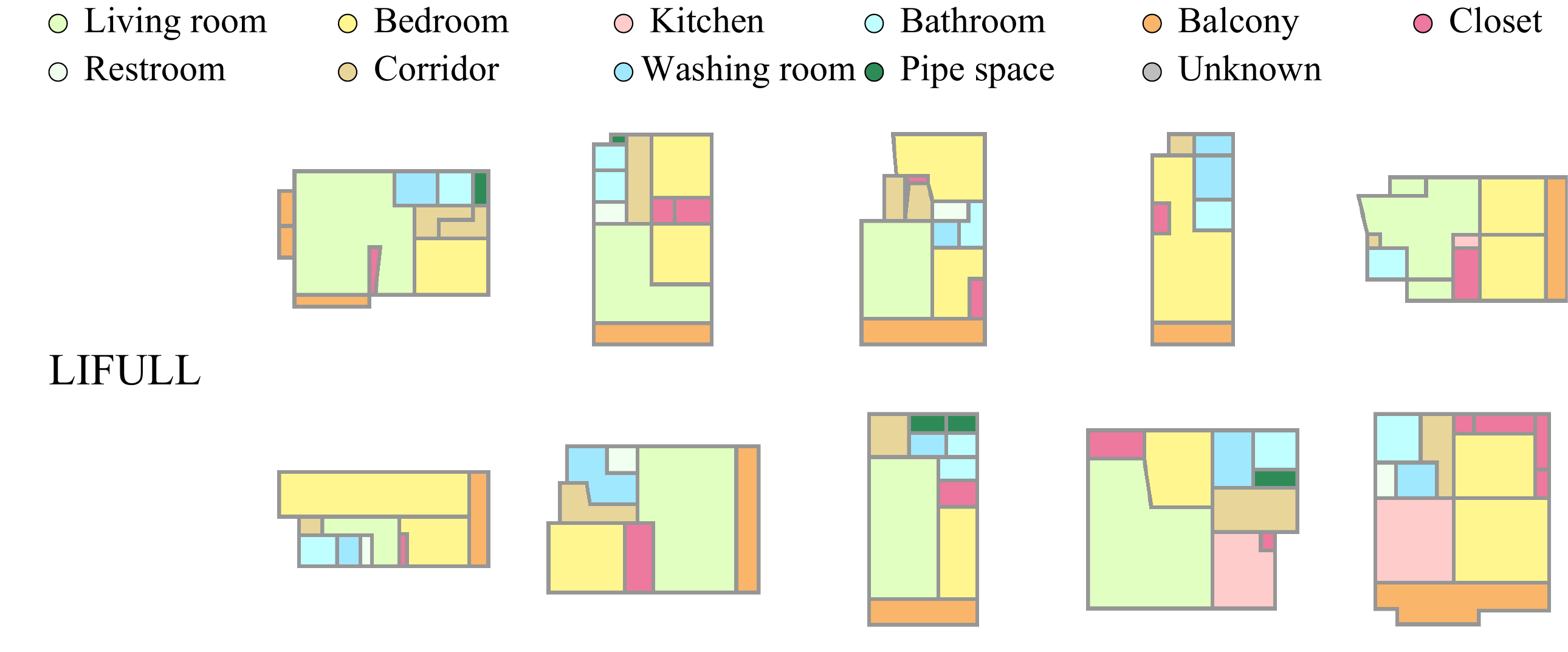}
	\caption{Qualitative evaluation on the \textit{LIFULL} dataset.}
	\label{fig:other-dataset}
\end{figure}

}

\section{Conclusion}

We introduce a novel vector floorplan generation framework,~\textit{GSDiff}, which converts the complex floorplan generation problem into a structural graph generation problem, further decoupled into node generation and edge prediction. Additionally, we incorporate geometric enhancements into the generation framework. By optimizing the node alignment error, we achieve better geometric consistency. Through edge perception enhancement, we improve the edge prediction ability, resulting in better geometric plausibility. Experiments have shown that~\textit{GSDiff} outperforms existing state-of-the-art methods.

However,~\textit{GSDiff} faces limitations, such as reliance on quality training data for learning complex node and edge dependencies and scalability challenges in larger projects. Future research directions include exploring semi-supervised learning to reduce data dependency, enhancing constraint handling for complex scenarios and improving scalability. These aim to broaden the framework's capabilities and applicability in architectural design.
\section*{Acknowledgments}

We would like to thank the anonymous reviewers for their constructive suggestions and comments.
This work is supported by the National Natural Science Foundation of China (62102126, 62372153) and the Fundamental Research Funds for the Central Universities of China (JZ2023HGTB0269, JZ2022HGQA0163). In this paper, we used "LIFULL HOME'S Dataset" provided by LIFULL Co., Ltd. via IDR Dataset Service of National Institute of Informatics.

\bibliography{aaai25}

\begin{thebibliography}{33}
\providecommand{\natexlab}[1]{#1}

\bibitem[{Abu-Aisheh et~al.(2015)Abu-Aisheh, Raveaux, Ramel, and
  Martineau}]{abu2015exact}
Abu-Aisheh, Z.; Raveaux, R.; Ramel, J.-Y.; and Martineau, P. 2015.
\newblock An exact graph edit distance algorithm for solving pattern
  recognition problems.
\newblock In \emph{4th International Conference on Pattern Recognition
  Applications and Methods 2015}.

\bibitem[{Bi{\'n}kowski et~al.(2018)Bi{\'n}kowski, Sutherland, Arbel, and
  Gretton}]{binkowski2018demystifying}
Bi{\'n}kowski, M.; Sutherland, D.~J.; Arbel, M.; and Gretton, A. 2018.
\newblock Demystifying mmd gans.
\newblock \emph{arXiv preprint arXiv:1801.01401}.

\bibitem[{Bisht et~al.(2022)Bisht, Shekhawat, Upasani, Jain, Tiwaskar, and
  Hebbar}]{bisht2022transforming}
Bisht, S.; Shekhawat, K.; Upasani, N.; Jain, R.~N.; Tiwaskar, R.~J.; and
  Hebbar, C. 2022.
\newblock Transforming an adjacency graph into dimensioned floorplan layouts.
\newblock In \emph{Computer Graphics Forum}, volume~41, 5--22. Wiley Online
  Library.

\bibitem[{Chaillou(2020)}]{chaillou2020archigan}
Chaillou, S. 2020.
\newblock Archigan: Artificial intelligence x architecture.
\newblock In \emph{Architectural intelligence: Selected papers from the 1st
  international conference on computational design and robotic fabrication
  (CDRF 2019)}, 117--127. Springer.

\bibitem[{Chen et~al.(2024)Chen, Zhang, Zhou, and Chen}]{chen2024towards}
Chen, J.; Zhang, R.; Zhou, Y.; and Chen, C. 2024.
\newblock Towards Aligned Layout Generation via Diffusion Model with Aesthetic
  Constraints.
\newblock \emph{arXiv preprint arXiv:2402.04754}.

\bibitem[{Heusel et~al.(2017)Heusel, Ramsauer, Unterthiner, Nessler, and
  Hochreiter}]{heusel2017gans}
Heusel, M.; Ramsauer, H.; Unterthiner, T.; Nessler, B.; and Hochreiter, S.
  2017.
\newblock Gans trained by a two time-scale update rule converge to a local nash
  equilibrium.
\newblock \emph{Advances in neural information processing systems}, 30.

\bibitem[{Ho, Jain, and Abbeel(2020)}]{ho2020denoising}
Ho, J.; Jain, A.; and Abbeel, P. 2020.
\newblock Denoising diffusion probabilistic models.
\newblock \emph{Advances in neural information processing systems}, 33:
  6840--6851.

\bibitem[{Ho et~al.(2022)Ho, Saharia, Chan, Fleet, Norouzi, and
  Salimans}]{ho2022cascaded}
Ho, J.; Saharia, C.; Chan, W.; Fleet, D.~J.; Norouzi, M.; and Salimans, T.
  2022.
\newblock Cascaded diffusion models for high fidelity image generation.
\newblock \emph{Journal of Machine Learning Research}, 23(47): 1--33.

\bibitem[{Hu et~al.(2020)Hu, Huang, Tang, Van~Kaick, Zhang, and
  Huang}]{hu2020graph2plan}
Hu, R.; Huang, Z.; Tang, Y.; Van~Kaick, O.; Zhang, H.; and Huang, H. 2020.
\newblock Graph2plan: Learning floorplan generation from layout graphs.
\newblock \emph{ACM Transactions on Graphics (TOG)}, 39(4): 118--1.

\bibitem[{Hu et~al.(2024)Hu, Wu, Su, Hou, Zheng, and Xu}]{raster-to-graph}
Hu, S.; Wu, W.; Su, R.; Hou, W.; Zheng, L.; and Xu, B. 2024.
\newblock Raster-to-Graph: Floorplan Recognition via Autoregressive Graph
  Prediction with an Attention Transformer.
\newblock \emph{Computer Graphics Forum}, 43(2): e15007.

\bibitem[{Kingma(2014)}]{kingma2014adam}
Kingma, D. 2014.
\newblock Adam: a method for stochastic optimization.
\newblock \emph{arXiv preprint arXiv:1412.6980}.

\bibitem[{Laignel et~al.(2021)Laignel, Pozin, Geffrier, Delevaux, Brun, and
  Dolla}]{laignel2021floor}
Laignel, G.; Pozin, N.; Geffrier, X.; Delevaux, L.; Brun, F.; and Dolla, B.
  2021.
\newblock Floor plan generation through a mixed constraint programming-genetic
  optimization approach.
\newblock \emph{Automation in Construction}, 123: 103491.

\bibitem[{LIFULL~Co.(2016)}]{lifull}
LIFULL~Co., L.~. 2016.
\newblock LIFULL HOME'S High Resolution Floor Plan Image Data. {Informatics
  Research Data Repository, National Institute of Informatics. (dataset).}
\newblock \url{https://doi.org/10.32130/idr.6.2}.

\bibitem[{Liu et~al.(2013)Liu, Yang, AlHalawani, and Mitra}]{liu2013constraint}
Liu, H.; Yang, Y.-L.; AlHalawani, S.; and Mitra, N.~J. 2013.
\newblock Constraint-aware interior layout exploration for pre-cast
  concrete-based buildings.
\newblock \emph{The Visual Computer}, 29: 663--673.

\bibitem[{Liu et~al.(2023)Liu, Wu, Van~Hoorick, Tokmakov, Zakharov, and
  Vondrick}]{liu2023zero}
Liu, R.; Wu, R.; Van~Hoorick, B.; Tokmakov, P.; Zakharov, S.; and Vondrick, C.
  2023.
\newblock Zero-1-to-3: Zero-shot one image to 3d object.
\newblock In \emph{Proceedings of the IEEE/CVF International Conference on
  Computer Vision}, 9298--9309.

\bibitem[{Merrell, Schkufza, and Koltun(2010)}]{merrell2010computer}
Merrell, P.; Schkufza, E.; and Koltun, V. 2010.
\newblock Computer-generated residential building layouts.
\newblock In \emph{ACM SIGGRAPH Asia 2010 papers}, 1--12.

\bibitem[{Nauata et~al.(2020)Nauata, Chang, Cheng, Mori, and
  Furukawa}]{nauata2020house}
Nauata, N.; Chang, K.-H.; Cheng, C.-Y.; Mori, G.; and Furukawa, Y. 2020.
\newblock House-gan: Relational generative adversarial networks for
  graph-constrained house layout generation.
\newblock In \emph{Computer Vision--ECCV 2020: 16th European Conference,
  Glasgow, UK, August 23--28, 2020, Proceedings, Part I 16}, 162--177.
  Springer.

\bibitem[{Nauata et~al.(2021)Nauata, Hosseini, Chang, Chu, Cheng, and
  Furukawa}]{nauata2021house}
Nauata, N.; Hosseini, S.; Chang, K.-H.; Chu, H.; Cheng, C.-Y.; and Furukawa, Y.
  2021.
\newblock House-gan++: Generative adversarial layout refinement network towards
  intelligent computational agent for professional architects.
\newblock In \emph{Proceedings of the IEEE/CVF Conference on Computer Vision
  and Pattern Recognition}, 13632--13641.

\bibitem[{Nichol et~al.(2021)Nichol, Dhariwal, Ramesh, Shyam, Mishkin, McGrew,
  Sutskever, and Chen}]{nichol2021glide}
Nichol, A.; Dhariwal, P.; Ramesh, A.; Shyam, P.; Mishkin, P.; McGrew, B.;
  Sutskever, I.; and Chen, M. 2021.
\newblock Glide: Towards photorealistic image generation and editing with
  text-guided diffusion models.
\newblock \emph{arXiv preprint arXiv:2112.10741}.

\bibitem[{Nichol et~al.(2022)Nichol, Jun, Dhariwal, Mishkin, and
  Chen}]{nichol2022point}
Nichol, A.; Jun, H.; Dhariwal, P.; Mishkin, P.; and Chen, M. 2022.
\newblock Point-e: A system for generating 3d point clouds from complex
  prompts.
\newblock \emph{arXiv preprint arXiv:2212.08751}.

\bibitem[{Para et~al.(2021)Para, Guerrero, Kelly, Guibas, and
  Wonka}]{para2021generative}
Para, W.; Guerrero, P.; Kelly, T.; Guibas, L.~J.; and Wonka, P. 2021.
\newblock Generative layout modeling using constraint graphs.
\newblock In \emph{Proceedings of the IEEE/CVF international conference on
  computer vision}, 6690--6700.

\bibitem[{Poole et~al.(2022)Poole, Jain, Barron, and
  Mildenhall}]{poole2022dreamfusion}
Poole, B.; Jain, A.; Barron, J.~T.; and Mildenhall, B. 2022.
\newblock Dreamfusion: Text-to-3d using 2d diffusion.
\newblock \emph{arXiv preprint arXiv:2209.14988}.

\bibitem[{Rombach et~al.(2022)Rombach, Blattmann, Lorenz, Esser, and
  Ommer}]{rombach2022high}
Rombach, R.; Blattmann, A.; Lorenz, D.; Esser, P.; and Ommer, B. 2022.
\newblock High-resolution image synthesis with latent diffusion models.
\newblock In \emph{Proceedings of the IEEE/CVF conference on computer vision
  and pattern recognition}, 10684--10695.

\bibitem[{Ronneberger, Fischer, and Brox(2015)}]{unet}
Ronneberger, O.; Fischer, P.; and Brox, T. 2015.
\newblock U-Net: Convolutional Networks for Biomedical Image Segmentation.
\newblock In Navab, N.; Hornegger, J.; Wells, W.~M.; and Frangi, A.~F., eds.,
  \emph{Medical Image Computing and Computer-Assisted Intervention -- MICCAI
  2015}, 234--241. Cham: Springer International Publishing.
\newblock ISBN 978-3-319-24574-4.

\bibitem[{Saharia et~al.(2022)Saharia, Chan, Chang, Lee, Ho, Salimans, Fleet,
  and Norouzi}]{saharia2022palette}
Saharia, C.; Chan, W.; Chang, H.; Lee, C.; Ho, J.; Salimans, T.; Fleet, D.; and
  Norouzi, M. 2022.
\newblock Palette: Image-to-image diffusion models.
\newblock In \emph{ACM SIGGRAPH 2022 conference proceedings}, 1--10.

\bibitem[{Shabani, Hosseini, and Furukawa(2023)}]{shabani2023housediffusion}
Shabani, M.~A.; Hosseini, S.; and Furukawa, Y. 2023.
\newblock Housediffusion: Vector floorplan generation via a diffusion model
  with discrete and continuous denoising.
\newblock In \emph{Proceedings of the IEEE/CVF Conference on Computer Vision
  and Pattern Recognition}, 5466--5475.

\bibitem[{Shekhawat et~al.(2021)Shekhawat, Upasani, Bisht, and
  Jain}]{shekhawat2021tool}
Shekhawat, K.; Upasani, N.; Bisht, S.; and Jain, R.~N. 2021.
\newblock A tool for computer-generated dimensioned floorplans based on given
  adjacencies.
\newblock \emph{Automation in Construction}, 127: 103718.

\bibitem[{Sun et~al.(2022)Sun, Wu, Liu, Min, Zhang, and
  Zheng}]{sun2022wallplan}
Sun, J.; Wu, W.; Liu, L.; Min, W.; Zhang, G.; and Zheng, L. 2022.
\newblock Wallplan: synthesizing floorplans by learning to generate wall
  graphs.
\newblock \emph{ACM Transactions on Graphics (TOG)}, 41(4): 1--14.

\bibitem[{van Engelenburg et~al.(2024)van Engelenburg, Mostafavi, Kuhn, Jeon,
  Franzen, Standfest, van Gemert, and Khademi}]{vanengelenburg2024msd}
van Engelenburg, C.; Mostafavi, F.; Kuhn, E.; Jeon, Y.; Franzen, M.; Standfest,
  M.; van Gemert, J.; and Khademi, S. 2024.
\newblock MSD: A Benchmark Dataset for Floor Plan Generation of Building
  Complexes.
\newblock arXiv:2407.10121.

\bibitem[{Vaswani et~al.(2017)Vaswani, Shazeer, Parmar, Uszkoreit, Jones,
  Gomez, Kaiser, and Polosukhin}]{vasvani2017attention}
Vaswani, A.; Shazeer, N.; Parmar, N.; Uszkoreit, J.; Jones, L.; Gomez, A.~N.;
  Kaiser, L.~u.; and Polosukhin, I. 2017.
\newblock Attention is All you Need.
\newblock In Guyon, I.; Luxburg, U.~V.; Bengio, S.; Wallach, H.; Fergus, R.;
  Vishwanathan, S.; and Garnett, R., eds., \emph{Advances in Neural Information
  Processing Systems}, volume~30. Curran Associates, Inc.

\bibitem[{Wang and Zhang(2020)}]{wang2020generating}
Wang, X.-Y.; and Zhang, K. 2020.
\newblock Generating layout designs from high-level specifications.
\newblock \emph{Automation in Construction}, 119: 103288.

\bibitem[{Wu et~al.(2018)Wu, Fan, Liu, and Wonka}]{wu2018miqp}
Wu, W.; Fan, L.; Liu, L.; and Wonka, P. 2018.
\newblock Miqp-based layout design for building interiors.
\newblock In \emph{Computer Graphics Forum}, volume~37, 511--521. Wiley Online
  Library.

\bibitem[{Wu et~al.(2019)Wu, Fu, Tang, Wang, Qi, and Liu}]{wu2019data}
Wu, W.; Fu, X.-M.; Tang, R.; Wang, Y.; Qi, Y.-H.; and Liu, L. 2019.
\newblock Data-driven interior plan generation for residential buildings.
\newblock \emph{ACM Transactions on Graphics (TOG)}, 38(6): 1--12.

\end{thebibliography}


\begin{thebibliography}{11}
\providecommand{\natexlab}[1]{#1}

\bibitem[{Chen et~al.(2024)Chen, Zhang, Zhou, and Chen}]{chen2024towards}
Chen, J.; Zhang, R.; Zhou, Y.; and Chen, C. 2024.
\newblock Towards Aligned Layout Generation via Diffusion Model with Aesthetic
  Constraints.
\newblock \emph{arXiv preprint arXiv:2402.04754}.

\bibitem[{Hu et~al.(2020)Hu, Huang, Tang, Van~Kaick, Zhang, and
  Huang}]{hu2020graph2plan}
Hu, R.; Huang, Z.; Tang, Y.; Van~Kaick, O.; Zhang, H.; and Huang, H. 2020.
\newblock Graph2plan: Learning floorplan generation from layout graphs.
\newblock \emph{ACM Transactions on Graphics (TOG)}, 39(4): 118--1.

\bibitem[{Kingma(2014)}]{kingma2014adam}
Kingma, D. 2014.
\newblock Adam: a method for stochastic optimization.
\newblock \emph{arXiv preprint arXiv:1412.6980}.

\bibitem[{Li et~al.(2020)Li, Yang, Zhang, Liu, Wang, and Xu}]{li2020attribute}
Li, J.; Yang, J.; Zhang, J.; Liu, C.; Wang, C.; and Xu, T. 2020.
\newblock Attribute-conditioned layout gan for automatic graphic design.
\newblock \emph{IEEE Transactions on Visualization and Computer Graphics},
  27(10): 4039--4048.

\bibitem[{Nauata et~al.(2021)Nauata, Hosseini, Chang, Chu, Cheng, and
  Furukawa}]{houseganpp}
Nauata, N.; Hosseini, S.; Chang, K.-H.; Chu, H.; Cheng, C.-Y.; and Furukawa, Y.
  2021.
\newblock House-gan++: Generative adversarial layout refinement network towards
  intelligent computational agent for professional architects.
\newblock In \emph{Proceedings of the IEEE/CVF Conference on Computer Vision
  and Pattern Recognition}, 13632--13641.

\bibitem[{Ronneberger, Fischer, and Brox(2015)}]{ronneberger2015u}
Ronneberger, O.; Fischer, P.; and Brox, T. 2015.
\newblock U-net: Convolutional networks for biomedical image segmentation.
\newblock In \emph{Medical image computing and computer-assisted
  intervention--MICCAI 2015: 18th international conference, Munich, Germany,
  October 5-9, 2015, proceedings, part III 18}, 234--241. Springer.

\bibitem[{Shabani, Hosseini, and Furukawa(2023)}]{shabani2023housediffusion}
Shabani, M.~A.; Hosseini, S.; and Furukawa, Y. 2023.
\newblock Housediffusion: Vector floorplan generation via a diffusion model
  with discrete and continuous denoising.
\newblock In \emph{Proceedings of the IEEE/CVF Conference on Computer Vision
  and Pattern Recognition}, 5466--5475.

\bibitem[{Sun et~al.(2022)Sun, Wu, Liu, Min, Zhang, and
  Zheng}]{sun2022wallplan}
Sun, J.; Wu, W.; Liu, L.; Min, W.; Zhang, G.; and Zheng, L. 2022.
\newblock Wallplan: synthesizing floorplans by learning to generate wall
  graphs.
\newblock \emph{ACM Transactions on Graphics (TOG)}, 41(4): 1--14.

\bibitem[{Vaswani et~al.(2017)Vaswani, Shazeer, Parmar, Uszkoreit, Jones,
  Gomez, Kaiser, and Polosukhin}]{vasvani2017attention}
Vaswani, A.; Shazeer, N.; Parmar, N.; Uszkoreit, J.; Jones, L.; Gomez, A.~N.;
  Kaiser, L.~u.; and Polosukhin, I. 2017.
\newblock Attention is All you Need.
\newblock In Guyon, I.; Luxburg, U.~V.; Bengio, S.; Wallach, H.; Fergus, R.;
  Vishwanathan, S.; and Garnett, R., eds., \emph{Advances in Neural Information
  Processing Systems}, volume~30. Curran Associates, Inc.

\bibitem[{Wu et~al.(2019)Wu, Fu, Tang, Wang, Qi, and Liu}]{wu2019data}
Wu, W.; Fu, X.-M.; Tang, R.; Wang, Y.; Qi, Y.-H.; and Liu, L. 2019.
\newblock Data-driven interior plan generation for residential buildings.
\newblock \emph{ACM Transactions on Graphics (TOG)}, 38(6): 1--12.

\bibitem[{Zhang et~al.(2017)Zhang, Cisse, Dauphin, and
  Lopez-Paz}]{zhang2017mixup}
Zhang, H.; Cisse, M.; Dauphin, Y.; and Lopez-Paz, D. 2017.
\newblock mixup: Beyond Empirical Risk Minimization.

\end{thebibliography}

\section{Reproducibility Checklist}

This paper:

Includes a conceptual outline and/or pseudocode description of AI methods introduced (yes)

Clearly delineates statements that are opinions, hypothesis, and speculation from objective facts and results (yes)

Provides well marked pedagogical references for less-familiare readers to gain background necessary to replicate the paper (yes)

\subsection{}

Does this paper make theoretical contributions? (no)

\subsection{}
Does this paper rely on one or more datasets? (yes)

If yes, please complete the list below.

A motivation is given for why the experiments are conducted on the selected datasets (yes)

All novel datasets introduced in this paper are included in a data appendix. (NA)

All novel datasets introduced in this paper will be made publicly available upon publication of the paper with a license that allows free usage for research purposes. (NA)

All datasets drawn from the existing literature (potentially including authors’ own previously published work) are accompanied by appropriate citations. (yes)

All datasets drawn from the existing literature (potentially including authors’ own previously published work) are publicly available. (yes)

All datasets that are not publicly available are described in detail, with explanation why publicly available alternatives are not scientifically satisficing. (NA)

\subsection{}
Does this paper include computational experiments? (yes)

If yes, please complete the list below.

Any code required for pre-processing data is included in the appendix. (no).

All source code required for conducting and analyzing the experiments is included in a code appendix. (no)

All source code required for conducting and analyzing the experiments will be made publicly available upon publication of the paper with a license that allows free usage for research purposes. (yes)

All source code implementing new methods have comments detailing the implementation, with references to the paper where each step comes from (partial)

If an algorithm depends on randomness, then the method used for setting seeds is described in a way sufficient to allow replication of results. (NA)

This paper specifies the computing infrastructure used for running experiments (hardware and software), including GPU/CPU models; amount of memory; operating system; names and versions of relevant software libraries and frameworks. (yes)

This paper formally describes evaluation metrics used and explains the motivation for choosing these metrics. (yes)

This paper states the number of algorithm runs used to compute each reported result. (yes)

Analysis of experiments goes beyond single-dimensional summaries of performance (e.g., average; median) to include measures of variation, confidence, or other distributional information. (yes)

The significance of any improvement or decrease in performance is judged using appropriate statistical tests (e.g., Wilcoxon signed-rank). (no)

This paper lists all final (hyper-)parameters used for each model/algorithm in the paper’s experiments. (yes)

This paper states the number and range of values tried per (hyper-) parameter during development of the paper, along with the criterion used for selecting the final parameter setting. (yes)

\end{document}


\maketitle

\section{Preliminary}

Diffusion models are a class of generative models that train neural networks to cleverly reverse a noise-adding process, enabling them to generate samples that simulate the original samples of the dataset from simple noise. A standard diffusion model typically consists of a forward process and a reverse process. 

In the forward process, noise is gradually added to the original sample \( V_0 \) over \( T \) steps, resulting in a noisy sample \( V_T \) that resembles Gaussian noise. For a given time step \( t \), this process generates a noisier version \( V_t \) using the following equation:
\begin{equation}
\label{1}
V_t = \sqrt{\overline{\alpha}_t} V_0 + \sqrt{1 - \overline{\alpha}_t} \epsilon
\end{equation}
where \( \overline{\alpha}_t = \alpha_1 \alpha_2 \cdots \alpha_t \) is the cumulative product of coefficients \( \alpha_i \) used to control the proportion of the signal \( V_0 \) in \( V_t \), and \( \epsilon \) is standard Gaussian noise. As \( t \) increases, the proportion of noise becomes larger until it approaches the pure Gaussian noise.

The reverse process starts with Gaussian noise and gradually denoises it to reconstruct the original sample \( V_0 \). Less noisy versions of the original sample are recovered step by step using a neural network parameterized by \( \theta \), which outputs \( \epsilon_\theta \) as an estimation of the added noise. The probability distribution at time \( t-1 \) for the noisy sample \( V_t \) is given by
\begin{equation}
\label{2}
p_\theta(V_{t-1} | V_t) = \mathcal{N}(V_{t-1}; \mu_\theta(V_t, t), \sigma^2 I)
\end{equation}
where
\begin{equation}
\label{3}
\mu_\theta(V_t, t) = \frac{1}{\sqrt{\alpha_t}} \left( V_t - \frac{1 - \alpha_t}{\sqrt{1 - \overline{\alpha}_t}} \epsilon_\theta(V_t, t) \right)
\end{equation}
\(\sigma^2\) is the variance associated with the diffusion process and \(\mathcal{N}\) denotes the Gaussian distribution.

\section{Node generation}

\subsection{Network architecture}

\begin{figure}[t]
	\includegraphics[width=0.75\linewidth]{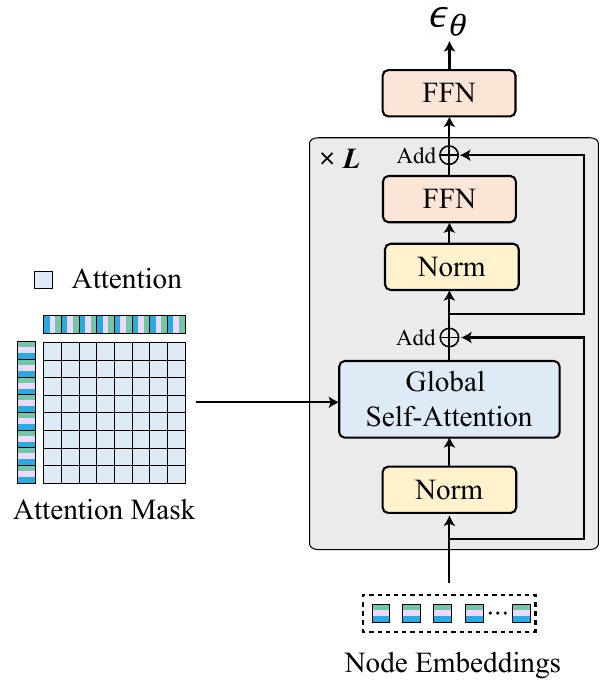}
	\centering
	\caption{The network architecture of the node Transformer.}
	\label{fig:node_network}
  \vspace{0mm}
\end{figure}

The network architecture of the node Transformer (Figure~\ref{fig:node_network}) consists of stacked Transformer decoder layers (×\( L \)).
The input to the node Transformer is the node set \( V_t \) and the time step \( t \). For a node \( v_i = (c_i, r_i, b_i) \in [-1, 1)^{2+7+1} \), we first obtain the node embedding \( f_i \in \mathbb{R}^d \), where \( d \) denotes the embedding dimension. Specifically, for the positional coordinates of the node, we first denormalize \( c_i = (x_i, y_i) \in [-1, 1)^2 \) back to \( [0, 255]^2 \) and use the positional encoding proposed by~\cite{vasvani2017attention}, \( f_i^c = [\gamma(x_i), \gamma(y_i)] \in \mathbb{R}^d \), where \( \gamma(t) \) is defined as
\begin{equation}
\label{4}
\gamma(t) = [\sin(\omega_0 t), \cos(\omega_0 t), \ldots, \sin(\omega_k t), \cos(\omega_k t)]
\end{equation}
where \( k = \frac{d}{4} - 1 \), \( \omega_j = \left(\frac{1}{10000}\right)^{\frac{4j}{d}} \), and \( d \) is the embedding dimension. For the semantic and background attributes \( (r_i, b_i) \) of the node, we use a fully connected layer to map \( (r_i, b_i) \) into a \( d \)-dimensional space to obtain the embedding \( f^{(r, b)}_i \in \mathbb{R}^d \). The time step \( t \) is encoded to represent the noise level, and we use FFN to obtain the time embedding \( f^t_i \in \mathbb{R}^d \). The final node embedding of \( v_i \) is obtained by fusing all three embeddings
\begin{equation}
\label{5}
f_i = f^{\text{c}}_i + f^{(r, b)}_i + f^t_i
\end{equation}
The network architecture of the node Transformer consists of multiple decoder layers, with the input being the node embeddings of all nodes. Each decoder layer includes normalization, multi-head global self-attention, residual connections, and FFN, where the attention module models the dependencies between all nodes. Considering the variable number of nodes of each sample, we use background nodes to pad the node set to the same shape. The output of the decoder is a set of node embeddings, which, after passing through several decoder layers, are used to predict the noise parameters \( \epsilon_\theta \) of \( p_\theta(V_{t-1} | V_t) \) at the next time step. 

\subsection{Alignment loss}

High-quality structural graphs of architectural floorplans require geometric consistency, meaning nodes must align well. The alignment loss is used to measure the alignment error between nodes of the structural graph. 
Inspired by~\cite{li2020attribute}, we define the alignment error of each node as the minimum distance between that node and all other nodes in any direction. Optimizing node alignment can be achieved by optimizing the sum of alignment errors of all nodes. We take the sum of alignment errors of all nodes as the alignment loss.

We use the predicted noise \( \epsilon_\theta \) and Equation~\ref{1} to predict node set \( \hat{V}_0 \). For \( \hat{V}_0 \), we have
\begin{equation}
\label{8}
\text{Alg}(\hat{V}_0) = \sum_{i=1}^n g(\min \left( \Delta b_i^X, \Delta b_i^Y \right))
\end{equation}
where \( n \) denotes the number of nodes in the node set, \( g(x) = -d \cdot \log(1 - \frac{x}{d}) \), \( \Delta b_i^* = \min_{\forall j \neq i} |b_i^* - b_j^*| \), \( * \in \mathcal{A} = \{X, Y\} \), and \( d \) represents the maximum allowable distance in direction \( * \).
Since the range of the positional coordinate for each node is \([-1, 1)\),  \( d = 1 - (-1) = 2 \).

However, the inherent regression errors of neural networks present a challenge. Directly optimizing the regression loss, typically the MSE loss) of the neural network does not work well. One solution is to use a discrete coordinate representation to optimize the alignment loss. \textit{HouseDiffusion}~\cite{shabani2023housediffusion} employs an 8-bit binary integer representation for coordinates within the range \([0, 255]\), which, in theory, could facilitate the model in learning precise alignment. However, this approach introduces notable risks in practice. For instance, consider a coordinate represented by the binary value $[1,0,0,0,0,0,1,1]^2$ (131). If a prediction error occurs at the 4\textsuperscript{th} bit, the actual predicted result would be $[1,0,0,0,1,0,1,1]^2$ (139), causing the coordinate to erroneously jump from 131 to 139. This magnitude of error significantly exceeds what would typically occur with direct regression methods.

%

To harmonize the advantages of binary representation and alignment loss, we convert the alignment error of each node into the binary form for regression. The discreteness introduced by the binary form helps to suppress noise to some extent, while the alignment loss constrains larger deviations. This method effectively avoids the pitfalls of high-bit inaccuracies and promotes coordinate alignment through discretization. The binary alignment loss is as follows
\begin{equation}
\label{9}
Alg^{2}(\hat{V}_0) = \sum_{i=1}^{n} g^{2}(\sum_{j=1}^{s} \left(Base^{2} \left( \min \left( \Delta b^X_i, \Delta b^Y_i \right) \right) \right)_j)
\end{equation}
where \(Base^{2}(\cdot)\) is the binary representation, \( \Delta b^*_i = \min_{j \neq i} | b^*_i - b^*_j | \), \( * \in \{X, Y\} \), \( g^{2}(x) = -d^{2} \log \left( 1 - \frac{x}{d^{2}} \right) \), with \( d^2 \) indicating the maximum allowable distance under the binary representation. \(n\) is the node number, \(s\) is the bit size. 
To calculate the binary alignment loss, we convert the coordinate range from \([-1, 1)\) to \([0, 255]\)  (corresponding to 8 bits). We take a precision of \(1 \times 10^{-1}\) (corresponding to 4-bit binary fraction), so \(s\) is set to 12 in our experiments. We use the L1 norm of the binary vector to represent the binary distance, therefore,
\begin{equation}
\label{91}
d^2 = \left\|\left[ \underbrace{1, 1, 1, \dots, 1, 1, 1}_{12} \right]^2 - \left[ \underbrace{0, 0, 0, \dots, 0, 0, 0}_{12} \right]^2 \right\|_1 = 12
\end{equation}
As we convert the coordinate range from \([-1, 1)\) to \([0, 255]\), we finally normalize the binary alignment loss by a scaling factor of \(1/128\). The notation \((\cdot)_j\) represents the \(j\)-th bit. 

\begin{figure}[t]
	\includegraphics[width=\linewidth]{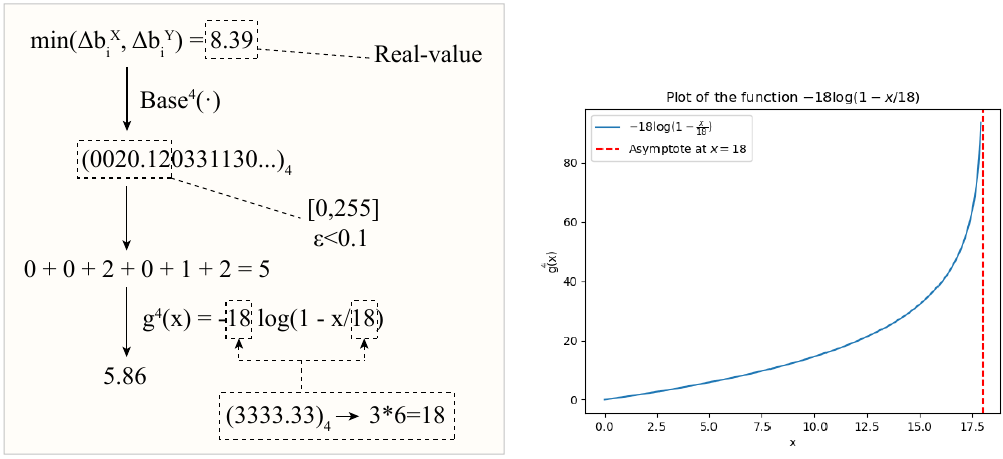}
	\centering
	\caption{Quaternary alignment error. \textbf{Left}: The calculation process of a quaternary alignment error. \textbf{Right}: Plot of \( g^{4}(x) = -18 \log \left( 1 - \frac{x}{18} \right) \). \( g^{4}(x)\) is used to impose penalties on larger alignment errors. When the independent variable (the alignment error of a single node) approaches 0, \( g^{4}(x)\) approximates \(x\); when the independent variable approaches \(x=d^4=18\), an infinite penalty will be imposed. The larger the independent variable, the greater the penalty.}
	\label{fig:alignment}
  \vspace{0mm}
\end{figure}

Binary learning treats high and low bits equally important, which is not optimal for gradient-based optimization. To balance precision and optimizability, we propose a mixed-base optimization strategy, combining multiple numeric bases: we trade off computational complexity (which increases linearly with the number of bases) and representation granularity by adding quaternary, octal, and hexadecimal bases to the binary and real number representations. 
The quaternary, octal, and hexadecimal losses are similar to those in binary. The bit size \( s \) is 6, 5, and 3 for these three bases, respectively. The L1 norm of the corresponding base vectors \( d^k \) is 18, 35, and 45, respectively. The normalization scaling factor is also \(1/128\). 
Without loss of generality, we illustrate the calculation process of a quaternary alignment error in Figure~\ref{fig:alignment}.

The final mixed-base alignment loss is given by
\begin{equation}
MixAlg(\hat{V}_0) = Alg(\hat{V}_0) + \sum_{k\in{2,4,8,16}} Alg^{k}(\hat{V}_0)
\end{equation}
\textit{LACE}~\cite{chen2024towards} suggests applying alignment loss at larger time \( t \), which may degrade performance, so they use a time-dependent constraint weight, applying larger alignment loss weights only at smaller time \( t \). We also apply this weight function, multiplying it with the alignment loss: \( \omega(t) = 1 - \alpha_{T-t} \) where \( \alpha_{T-t} \) is from Equation~\ref{1}. 



\subsection{Clamping}

During sampling (i.e., the reverse process of the diffusion model), the neural network's estimation of data at \( t = 0 \) may exceed reasonable ranges, causing deviation in the sampling path and degrading generation quality. The clamping technique limits the neural network's output to reasonable ranges to prevent such occurrences. In our case, the normalized coordinates for node attributes are limited to \( [-1, 1)^2 \), and other attributes are constrained to \(\{0, 1\}\). According to Equation~\ref{2}, the predicted noise \( \epsilon_\theta \) in the reverse process will be converted into the prediction \( \hat{V}_0 \). We clamp \( \hat{V}_0 \) within these ranges at each step of the reverse process to improve generation quality.
We set the threshold of all semantic attributes to 0.5 and the threshold of background attributes to 0.75 to avoid missing nodes.

\section{Edge prediction}

\subsection{Network architecture}

\begin{figure}[t]
	\includegraphics[width=0.75\linewidth]{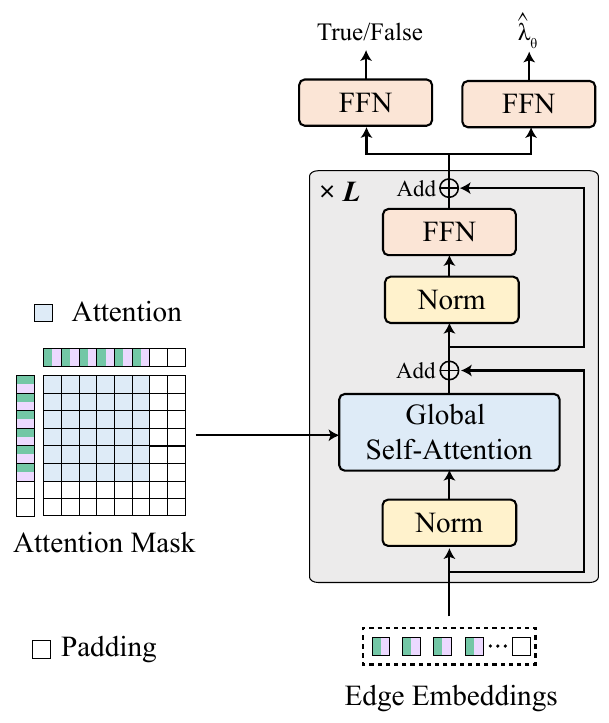}
	\centering
	\caption{The network architecture of the edge Transformer.}
	\label{fig:edge}
  \vspace{0mm}
\end{figure}

We use the edge Transformer to perform binary classification (``True/False") on the candidate edges. The network architecture of the edge Transformer (Figure~\ref{fig:edge}) consists of stacked Transformer decoder layers (×\( L \)), each comprising normalization, multi-head global self-attention, residual connections, and FFN. The input to the Edge Transformer consists of the embeddings of the candidate edge set \( E' \) and the padding mask constructed for \( E' \), where an edge is considered non-padding only if both of its endpoints are non-padding. 
After passing through several decoder layers, the output embeddings of the candidate edges are fed through an FFN to predict the authenticity of the edges. All edges identified as ``True" form the edge set \( E \), which is the output of the edge prediction.

\subsection{Edge perception enhancement}

\begin{figure}[t]
	\includegraphics[width=0.75\linewidth]{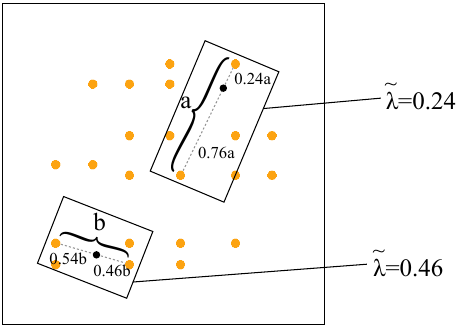}
	\centering
	\caption{Edge perception enhancement. This figure shows the relationship between \(\tilde{\lambda}\) and the position of the interpolation point for candidate edges of lengths a and b. For the candidate edge of length a, the interpolation point divides the candidate edge into two segments with a length ratio of 0.76:0.24 (regardless of whether the interpolation coefficient \(\lambda\) is 0.24 or 0.76), then \(\tilde{\lambda}\) is equal to the smaller of the two, which is 0.24. The interpolation coefficient \(\lambda\) is random each time, ensuring that every point on the entire candidate edge can be selected. }
	\label{fig:perception}
  \vspace{0mm}
\end{figure}


In general, the edge prediction model only uses the features of two endpoints as the edge features, which lack sufficient geometric information, resulting in missing or false edges and making the edge prediction unreasonable. 
To enhance the ability of edge perception, we propose a self-supervised edge perception enhancement strategy, as shown in Figure~\ref{fig:perception}. Specifically, for each candidate edge, we introduce a third point at a random location on the edge, determined by a random coefficient \(\lambda\). 
As a result, each edge is no longer represented solely by two endpoints, but by two endpoints along with a randomly interpolated point. The model is trained to predict this interpolation coefficient \(\lambda\), which forces the model to better perceive edges. This strategy improves the geometric reasoning capability of the model by introducing additional geometric features for each edge.

For each candidate edge \((v_i, v_j)\), the enhanced edge features include the features of both endpoints, as well as the feature combination of the random interpolation point. The interpolation point feature is determined by the interpolation coefficient \(\lambda\), which decides the location of the interpolation point on the edge. The coordinates and semantics of the interpolation point are as follows
\begin{equation}
c_\lambda = \lambda c_i + (1 - \lambda) c_j \in [-1, 1)^2
\end{equation}
\begin{equation}
r_\lambda = \textbf{0} \in \mathbb{R}^7
\end{equation}
where \(\lambda \sim U(0, 1)\) is a uniformly distributed interpolation coefficient sampled from the range \([0, 1]\). The zero vector \(\textbf{0}\) represents semantics, as the semantics at the random interpolation point on any candidate edge do not contribute to geometric reasoning. The loss for the self-supervised term is defined as
\begin{equation}
\tilde{\lambda} = 
\begin{cases} 
1 - \lambda, & \text{if } \lambda > 0.5, \\
\lambda, & \text{otherwise}.
\end{cases}
\end{equation}
\begin{equation}
\mathcal{L}_\lambda = \mathbb{E} \left[ \left| \tilde{\lambda} - \hat{\lambda}_\theta(e_{ij}) \right| \right]
\end{equation}
where \(\hat{\lambda}_\theta(e_{ij})\) represents the predicted interpolation coefficient. We process the original interpolation coefficient \(\lambda \in [0, 1]\) as follows: if \(\lambda > 0.5\), we use \(1 - \lambda\); otherwise, we keep it as is. This is done because the edges are undirected, so the model cannot distinguish which endpoint is the reference point during prediction. The total loss is defined as
\begin{equation}
\mathcal{L}_{\text{edge}} = \mathcal{L}_{\text{cls}} + \mathcal{L}_\lambda
\end{equation}
where \(\mathcal{L}_{cls} \) is the Cross-entropy classification loss, and \(\mathcal{L}_\lambda\) is the interpolation coefficient regression loss.

It is noteworthy that our edge perception enhancement strategy differs significantly from the mixup technique~\cite{zhang2017mixup} in both implementation and objective. The mixup technique creates new samples by blending two random samples from the training data, aiming to improve the model's generalization and its resistance to adversarial samples. In contrast, our strategy is tailored to enhance the model's perception of the geometric characteristics of edges within a structure. We assign a distinct interpolation coefficient to each candidate edge and train the model to predict this coefficient, enabling the model to discern the continuity of points along the edge, treating the candidate edge as a whole line segment rather than just two endpoints. This is crucial for the edge prediction task, which involves geometric inference. Unlike the mixup technique, whose goal is to smooth the decision boundary in the model's output space, our strategy focuses on enriching the input feature space with geometric features that are directly pertinent to the current task.

\subsection{Floorplan extraction}

\begin{figure}[t]
	\includegraphics[width=0.9\linewidth]{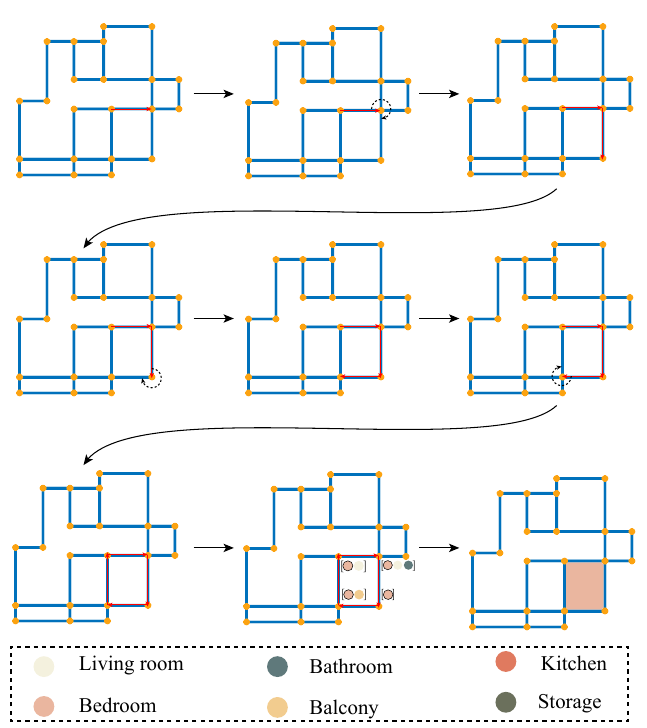}
	\centering
	\caption{An example of floorplan extraction.}
	\label{fig:extraction}
  \vspace{0mm}
\end{figure}

The floorplan extraction process is easy to implement.
We can simply extract all minimal polygonal cycles as rooms. This process involves starting from a certain edge in the structural graph, following a predefined order of nodes, moving from the smaller-numbered endpoint to the larger-numbered endpoint, and consistently turning the maximum angle in a fixed clockwise (or counterclockwise) direction to select the corresponding next edge. This process continues until it returns to the starting edge, and all edges traversed during this process form a minimal polygonal cycle. This procedure is illustrated in Figure~\ref{fig:extraction}.
Specifically, we traverse all edges to obtain all polygonal cycles and select the non-duplicate ones. In practice, we apply the following simple rule to avoid obtaining duplicate polygons: for each edge, if it has been traversed in the order from the smaller-numbered endpoint to the larger-numbered endpoint, it is marked as visited and will not be visited again. Each time, we only traverse edges that have not been visited.

\section{Constrained generation}

\subsection{Training paradigm}

We do not directly train the constraint encoder and node generation model jointly, as this would limit training on a training dataset of finite size. Given the critical importance of the quality of the constraint encoder, we first train each constraint encoder separately to achieve near-lossless performance, and then train the constraint encoder and node generation model jointly. To train the constraint encoder, we adopt a pre-training + fine-tuning paradigm: we first perform pre-training using randomly generated samples (which can be considered as having an infinite size of the training dataset), and then fine-tune the model on our training dataset. This strategy enhances generalization capability. The training is based on the reconstruction loss of the autoencoder.

\subsection{Boundary-constrained generation} 

\begin{figure}[t]
	\includegraphics[width=\linewidth]{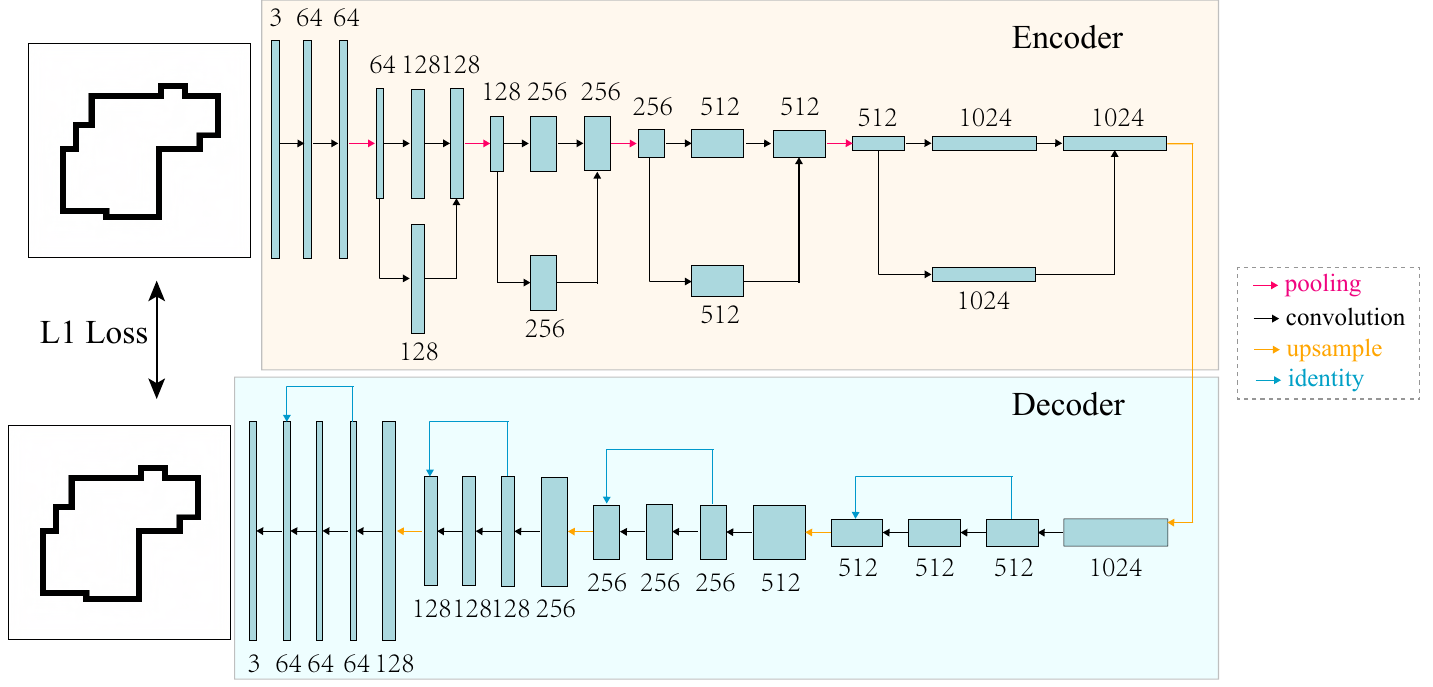}
	\centering
	\caption{The network architecture of the autoencoder for boundary constraints.}
	\label{fig:boundary}
  \vspace{0mm}
\end{figure}

In architectural floorplans, the boundary refers to the contour formed by the outer walls of the building, typically represented as a polygon. Given its geometric nature, we use the Convolutional Neural Network (CNN) to encode it. Specifically, during the pre-training phase, the input to the CNN is heuristically constructed. We determine each polygon vertex through a random walk on a \( 256 \times 256 \) three-channel blank image, where each vertex's 2D coordinates are uniformly sampled across the entire image. The edges of the polygon are drawn in layers with different colors: green (7 pixels), blue (5 pixels), red (3 pixels), and black (1 pixel). The number of vertices, corresponding to the number of random walk steps, is sampled from the training dataset.

We modified the U-Net structure~\cite{ronneberger2015u} by removing the skip connections and converting it into an autoencoder, adding residual connections between layers to improve performance. As the network depth increases, the number of channels in the encoder gradually increases, and we replace identity mapping with \( 1 \times 1 \) convolution. The network output is also a \( 256 \times 256 \) three-channel image, and we compute the L1 loss between the input and output images. Figure~\ref{fig:boundary} illustrates the network architecture of the autoencoder for boundary constraints.
For fine-tuning, we use the real boundary sample from the training dataset. The boundaries are drawn with black lines 7 pixels wide, and the same loss is used for training.

\subsection{Topology-constrained generation} 

The topological graph is defined as an undirected graph \( G_{\text{top}} = (V_{\text{top}}, E_{\text{top}}) \), where each vertex \( v_{\text{top}, i} = (r_i) \in V_{\text{top}} \) represents a room, and a one-hot encoded vector \( r_i \in \{0, 1\}^7 \) indicates the room category. The edge \( (v_{\text{top}, i}, v_{\text{top}, j}) \in E_{\text{top}} \) represents an adjacency relationship between a pair of rooms (i.e., whether they share a wall). Considering the graph-like nature of this problem, we use the topology Transformer to encode it. 

In the pre-training, we pre-compute room counts (ranging from 4 to 8), adjacency relationships (``True" or ``False"), and room categories from the dataset. We then randomly sample the room count, adjacency relationships, and room categories to construct random topological graphs, which are used as input to the topology Transformer. Since the room count in the topological graph is uncertain, padding is required, and a cross-attention mask is used to limit attention to all nodes and real rooms. The input to the topology Transformer is \( G_{\text{top}} \) and the cross-attention mask, where the room set \( V_{\text{top}} \) attends to the given adjacency relationships \( E_{\text{top}} \). After passing through several encoder layers, the output is the room embeddings.

\begin{figure}[t]
	\includegraphics[width=0.75\linewidth]{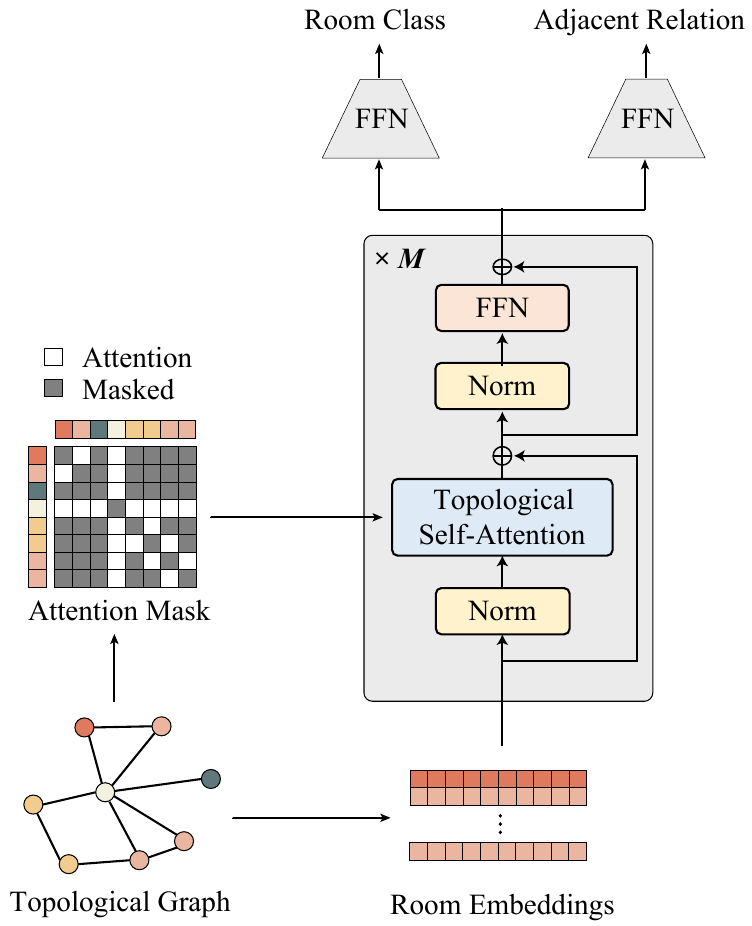}
	\centering
	\caption{The network architecture of the topology Transformer. \textcolor[rgb]{0.5,0.5, 0.5}{Gray} in attention mask: which attention are masked.}
	\label{fig:topology}
  \vspace{0mm}
\end{figure}

In the fine-tuning, we perform room classification and adjacency relationship classification tasks on the topological graph as reconstruction tasks: room classification forces room embeddings to contain their category information, while adjacency relationship classification forces room embeddings to contain the correct adjacency relationships. As a result, room embeddings effectively encode the entire information of the topological graph, serving as a constraint for node generation. Both room classification and adjacency relationship classification use the cross-entropy loss. The network architecture of the topology Transformer is shown in Figure~\ref{fig:topology}.
We fine-tune using the real topological graphs from the training dataset.

\section{Experimental setup}

The implementation details of the different networks are as follows, with hyperparameters chosen based on empirical observations and performance on the validation dataset.

\subsection{Unconstrained generation}

\paragraph{Node Transformer} The node Transformer consists of 24 layers with an embedding dimension of 256, resulting in a total of 19 million parameters. The batch size is set to 256, and the optimizer is Adam~\cite{kingma2014adam}. The training is conducted for 1,000,000 steps. The initial learning rate is \(1 \times 10^{-4}\), which is reduced by a factor of 0.1 after 500,000 steps.

\paragraph{Edge Transformer} The edge Transformer consists of 12 layers with an embedding dimension of 256, resulting in a total of 10 million parameters. The batch size is set to 8. Given the significant impact of edge prediction quality on the results, we implement a learning rate decay strategy for training: the initial learning rate is \(1 \times 10^{-4}\), and the optimizer is Adam~\cite{kingma2014adam}. The performance on the validation dataset is closely monitored, and if the validation metric shows no improvement for five consecutive evaluations, the learning rate is reduced by a factor of 0.1. If there is no improvement for 20 consecutive evaluations, training is terminated, and the model with the best performance on the validation dataset is selected. Evaluation is performed every 1,000 steps. The performance of the edge prediction model peaked at 61,000 steps.

\subsection{Constrained generation}
For the constrained generation, we increased the embedding dimension of the node Transformer to 512, resulting in a total of 96 million parameters. This change is made as we have observed that using the same 19 million parameters, adding topological constraints could reduce performance, possibly due to the difficulty of accommodating multiple types of information such as topological graphs and coordinates within the 256-dimensional space. The boundary constraints are configured similarly. The configuration of the Edge Transformer remained unchanged.

\paragraph{Boundary CNN} The boundary CNN has 31 million parameters. During the pre-training phase, the batch size is set to 16, and the optimizer is Adam~\cite{kingma2014adam} with an initial learning rate of \(1 \times 10^{-4}\). We use the same learning rate decay strategy for training as the edge Transformer. The performance peaked at 5,000 steps during pre-training. For fine-tuning, we restarted the learning rate at \(1 \times 10^{-4}\) with a batch size of 16, and evaluation was performed every 100 steps, continuing training until 6,700 steps.

\paragraph{Topology Transformer} The topology Transformer consists of 24 layers with an embedding dimension of 256, resulting in a total of 19 million parameters. During the pre-training phase, the batch size is set to 2048, and the optimizer is Adam~\cite{kingma2014adam} with an initial learning rate of \(1 \times 10^{-4}\). We use the same learning rate decay strategy for training as the edge Transformer. The performance peaked at 11,000 steps during pre-training. For fine-tuning, we restarted the learning rate at \(1 \times 10^{-4}\) with a batch size of 256, continuing training for 6,000 steps with the same method.

\section{Dataset}

\begin{figure*}[t]
	\includegraphics[width=\linewidth]{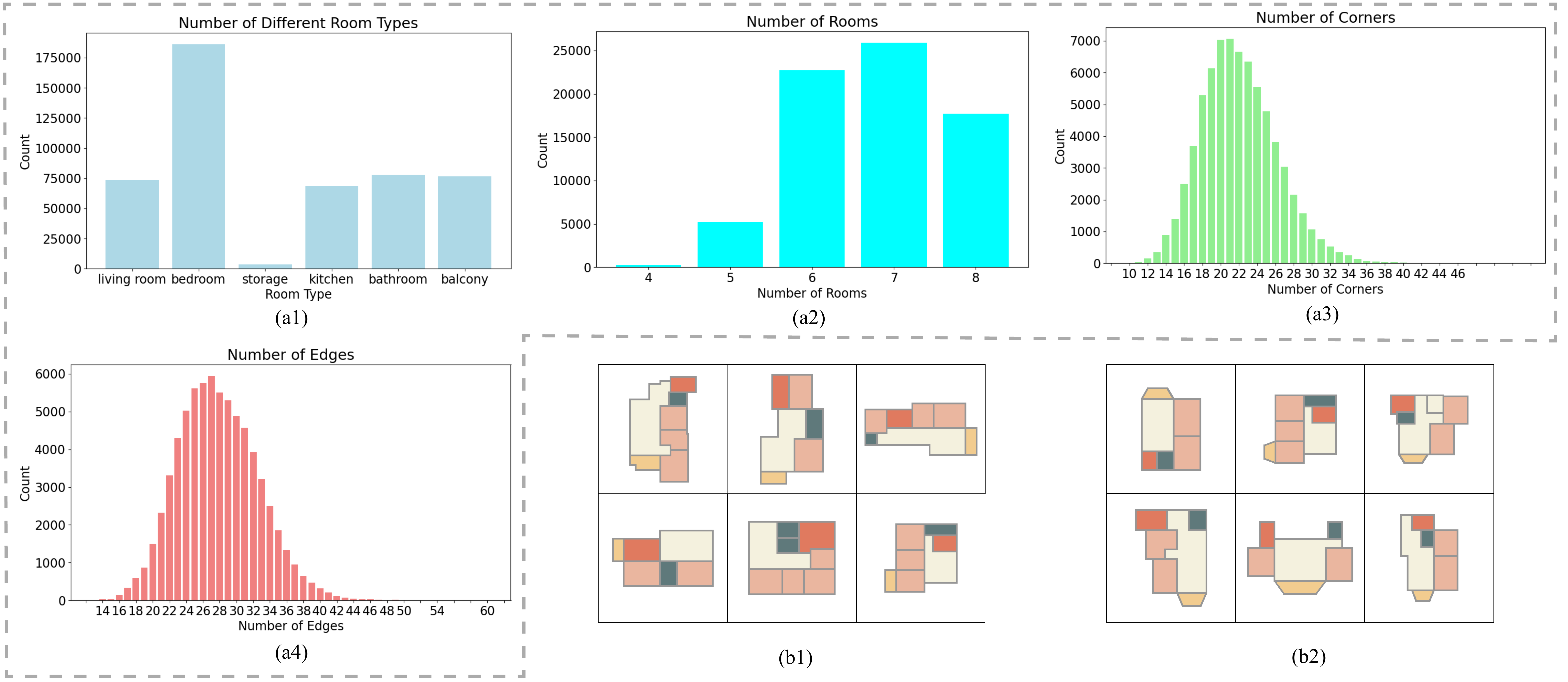}
	\centering
	\caption{Dataset. (a1)-(a4): distributional statistics on the processed RPLAN dataset~\cite{wu2019data}, including the quantity of each room category (a1), the number of rooms (a2), the number of wall junctions (a3) and the number of wall segments (a4). (b1)-(b2): some data samples obtained from the above data processing, (b1) shows original floorplans, and (b2) shows floorplans with slanted walls.}
	\label{fig:dataset}
  \vspace{0mm}
\end{figure*}

We first extract the 2D coordinates of wall junctions and segments. The image is binarized, with the walls represented as white and all other areas as black. Due to varying wall thicknesses, we regularize the thickness through repeated morphological operations (erosion, dilation) and template matching. We iteratively erode the white pixels representing the wall junction components until the next erosion step results in a decrease in the number of connected components in the image. This indicates that some wall segments have been eroded to a thickness of 1 pixel.

Next, we apply a series of template matching operations, sliding a 3×3 window across the image at this stage. If a match is successful, the matching area (the 3×3 window) is marked as white. The templates represent lines with a thickness of 1 pixel or local wall shapes that are defective. Through this process, the shape of the walls is gradually standardized, and the thickness becomes more uniform. This iterative process continues until the erosion reduces the thickness of all walls to 1 pixel; at this point, further erosion would cause all walls to disappear, turning the image completely black. At this stage, the wall structure is what we require.

Finally, the wall junctions and segments are extracted from the image, where the wall thickness has been uniformly reduced to 1 pixel. We obtain semantics from the four-channel images of the original RPLAN dataset. Any images that fail to process at this step are discarded. We obtain 71,763 vectorized floorplan images, randomly splitting them into 3,000 for the validation set, 3,000 for the test set, and the remainder for the training set. In the original RPLAN dataset, rooms are divided into 14 categories, which we merged into 7 categories: \textit{Living room}, \textit{Bedroom}, \textit{Kitchen}, \textit{Bathroom}, \textit{Balcony}, \textit{Storage}, \textit{External area}. 

Figure~\ref{fig:dataset} (b1) shows some floorplan samples obtained from the above process. The RPLAN dataset~\cite{wu2019data} comes from real residential layouts, which do not contain slanted walls. To verify that our method is also applicable to floorplans with slanted walls, we heuristically deform the ``peninsula-like" rectangular balcony which surrounded by the \textit{External area} on three sides into an isosceles trapezoid with the top base being 0.618 times the length of the bottom base. The data after the slanting deformation is shown in Figure~\ref{fig:dataset} (b2).

We have conducted distributional statistics (Figure~\ref{fig:dataset}(a)) on the processed RPLAN dataset~\cite{wu2019data}, including the number of wall junctions, the number of wall segments, the number of rooms, and the quantity of each room category. 

\section{More results}
\begin{figure*}[t]
	\includegraphics[width=\linewidth]{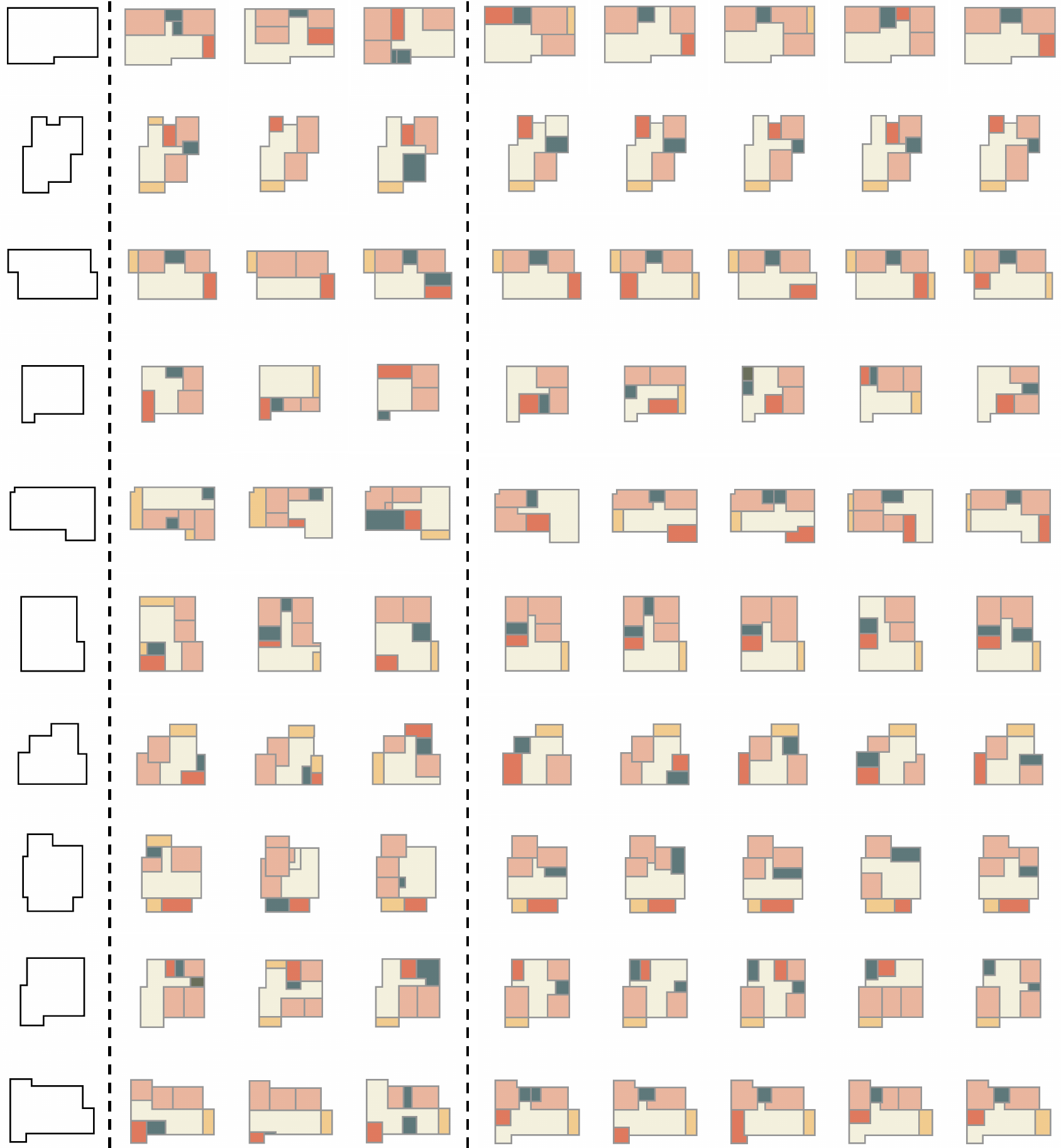}
	\put(-510, -15){Input boundary}
	\put(-430, -15){GT}
	\put(-390, -15){\textit{Graph2Plan}}
	\put(-328, -15){\textit{WallPlan}}
	\put(-148, -15){Ours}
	\centering
     \vspace{0mm}
	\caption{More on the comparison of boundary-constrained generation across different methods.}
	\label{fig:results1}
  \vspace{0mm}
\end{figure*}

\begin{figure*}[t]
	\includegraphics[width=\linewidth]{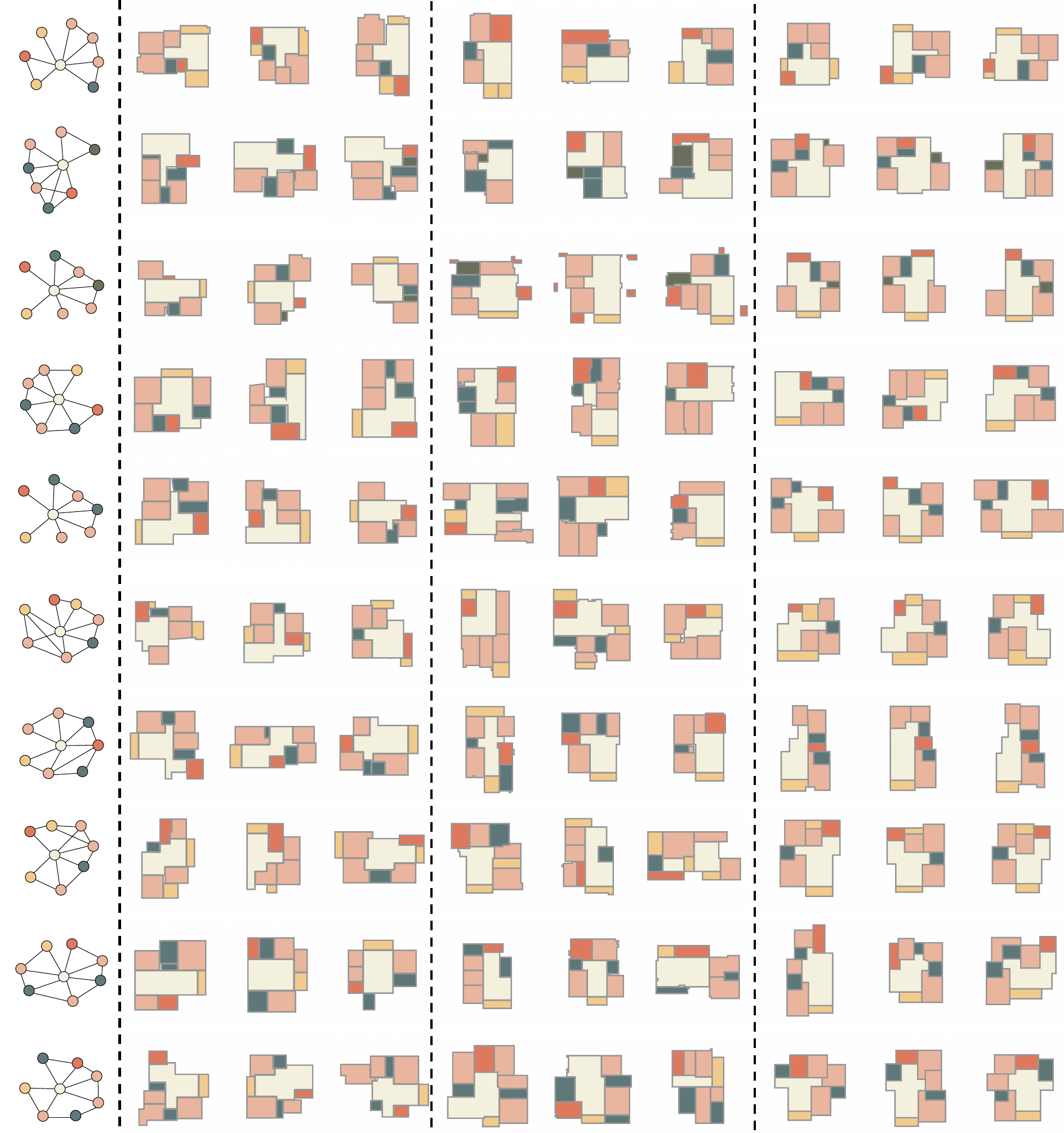}
	\put(-505, -15){Input topology}
	\put(-400, -15){\textit{HouseDiffsion}}
	\put(-250, -15){\textit{House-GAN++}}
	\put(-80, -15){Ours}
	\centering
	\caption{More on the comparison of topology-constrained generation among various techniques.}
	\label{fig:results2}
  \vspace{0mm}
\end{figure*}

\begin{figure*}[t]
	\includegraphics[width=\linewidth]{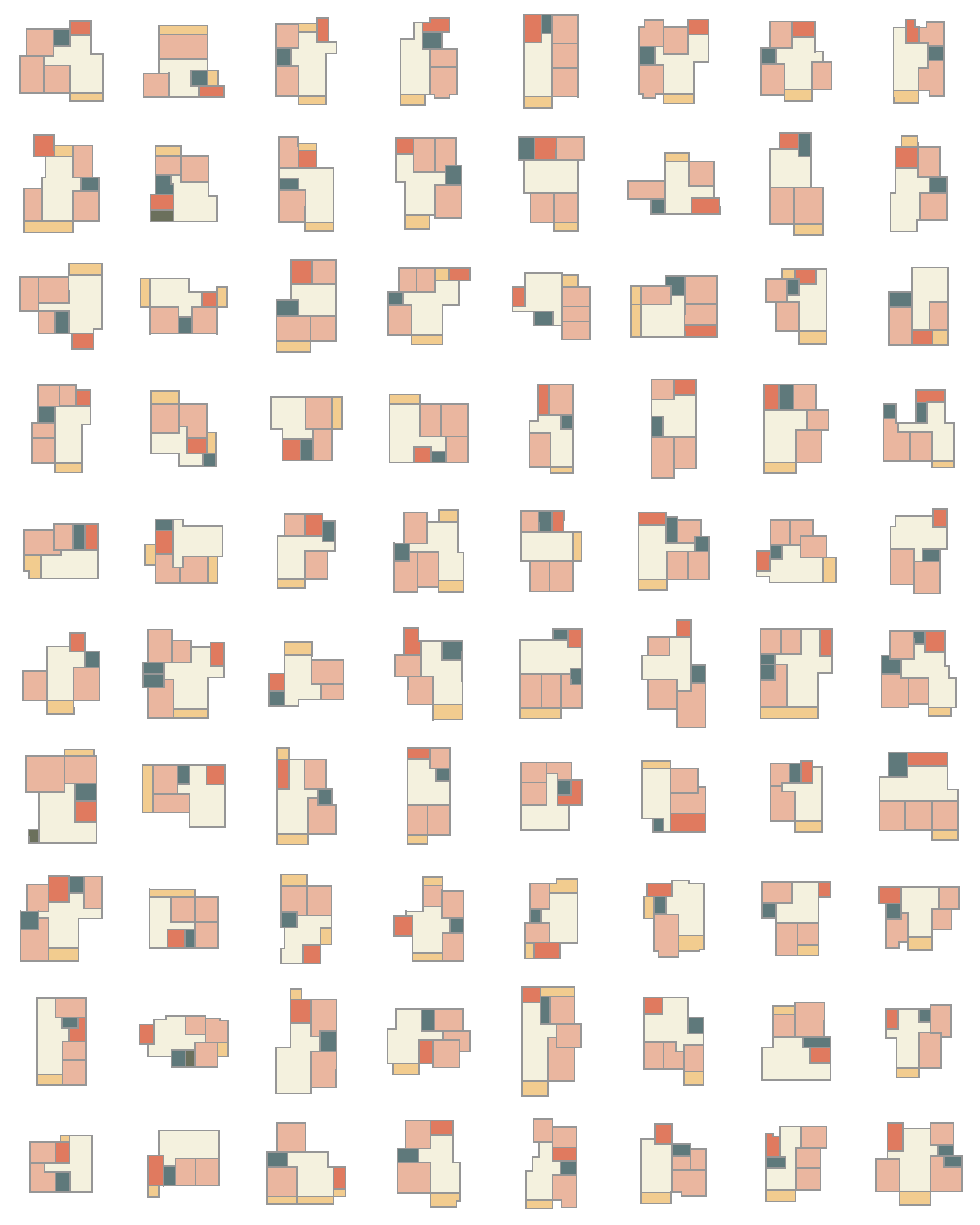}
	\centering
	\caption{More results of unconstrained generation by our method.}
	\label{fig:results3}
  \vspace{0mm}
\end{figure*}

\begin{figure*}[t]
	\includegraphics[width=\linewidth]{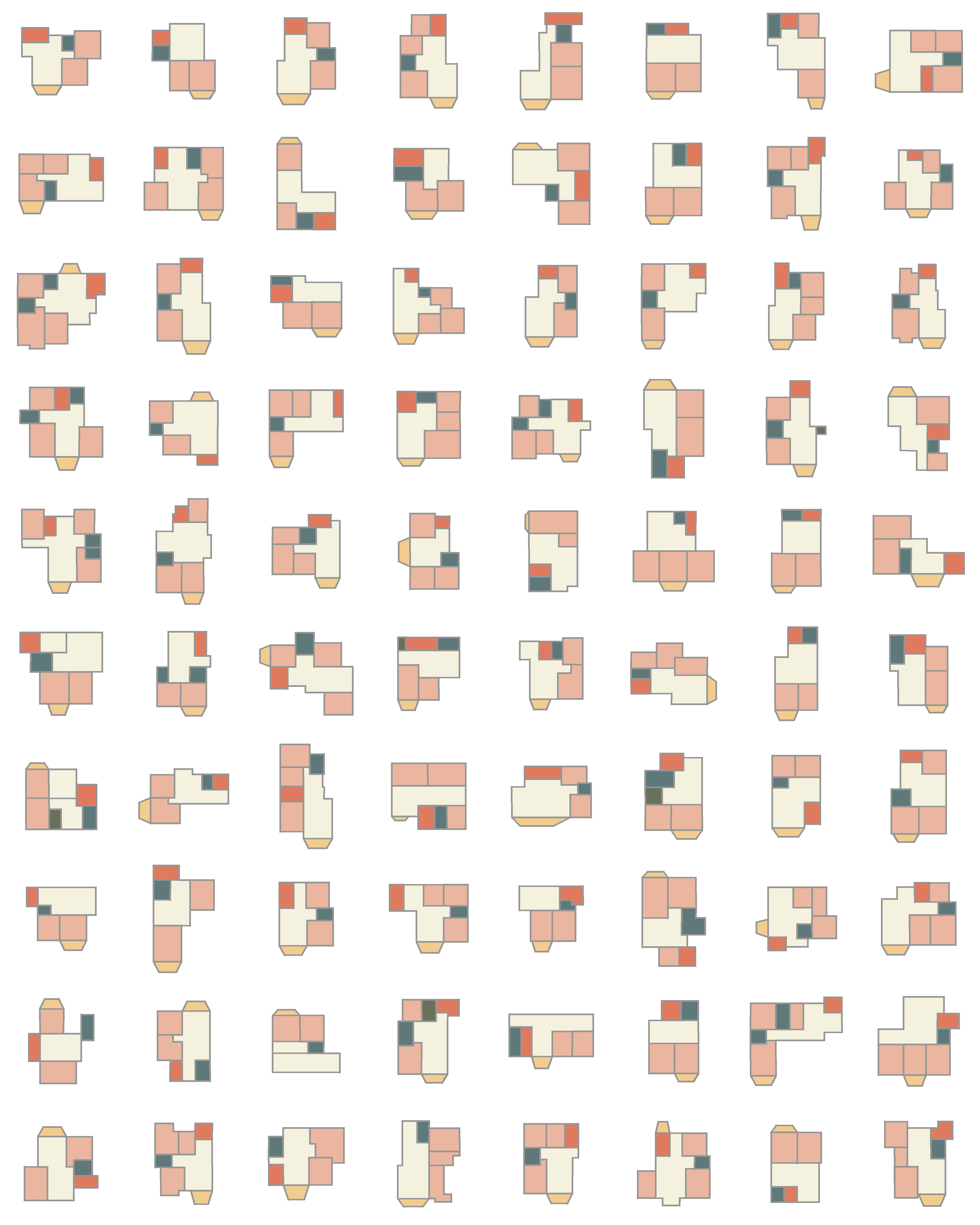}
	\centering
	\caption{More results of unconstrained generation with slanted walls by our method.}
	\label{fig:results4}
  \vspace{0mm}
\end{figure*}

\begin{figure*}[t]
	\includegraphics[width=\linewidth]{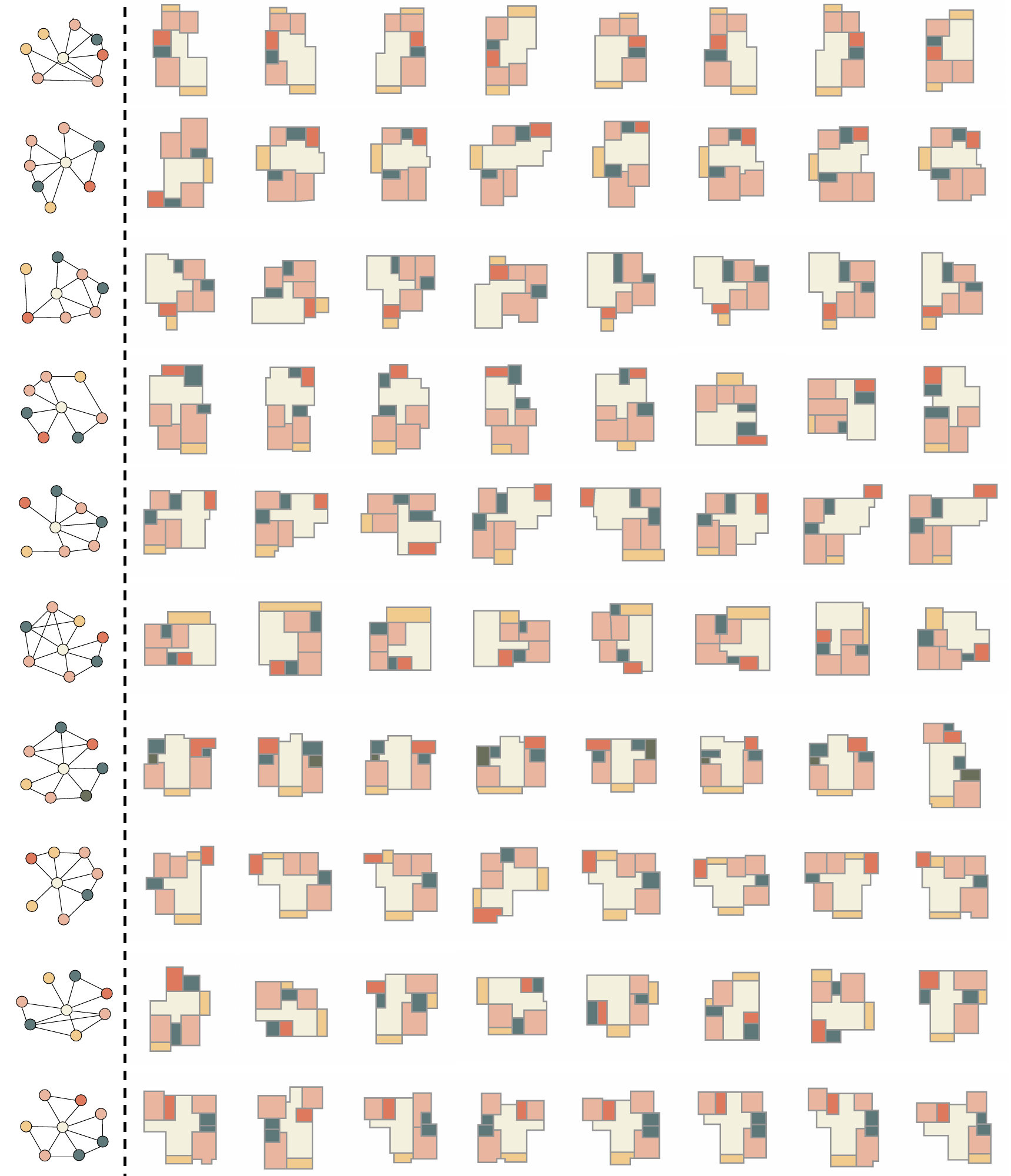}
	\centering
	\caption{More results of topology-constrained generation by our method: part (I).}
	\label{fig:results7}
  \vspace{0mm}
\end{figure*}

\begin{figure*}[t]
	\includegraphics[width=\linewidth]{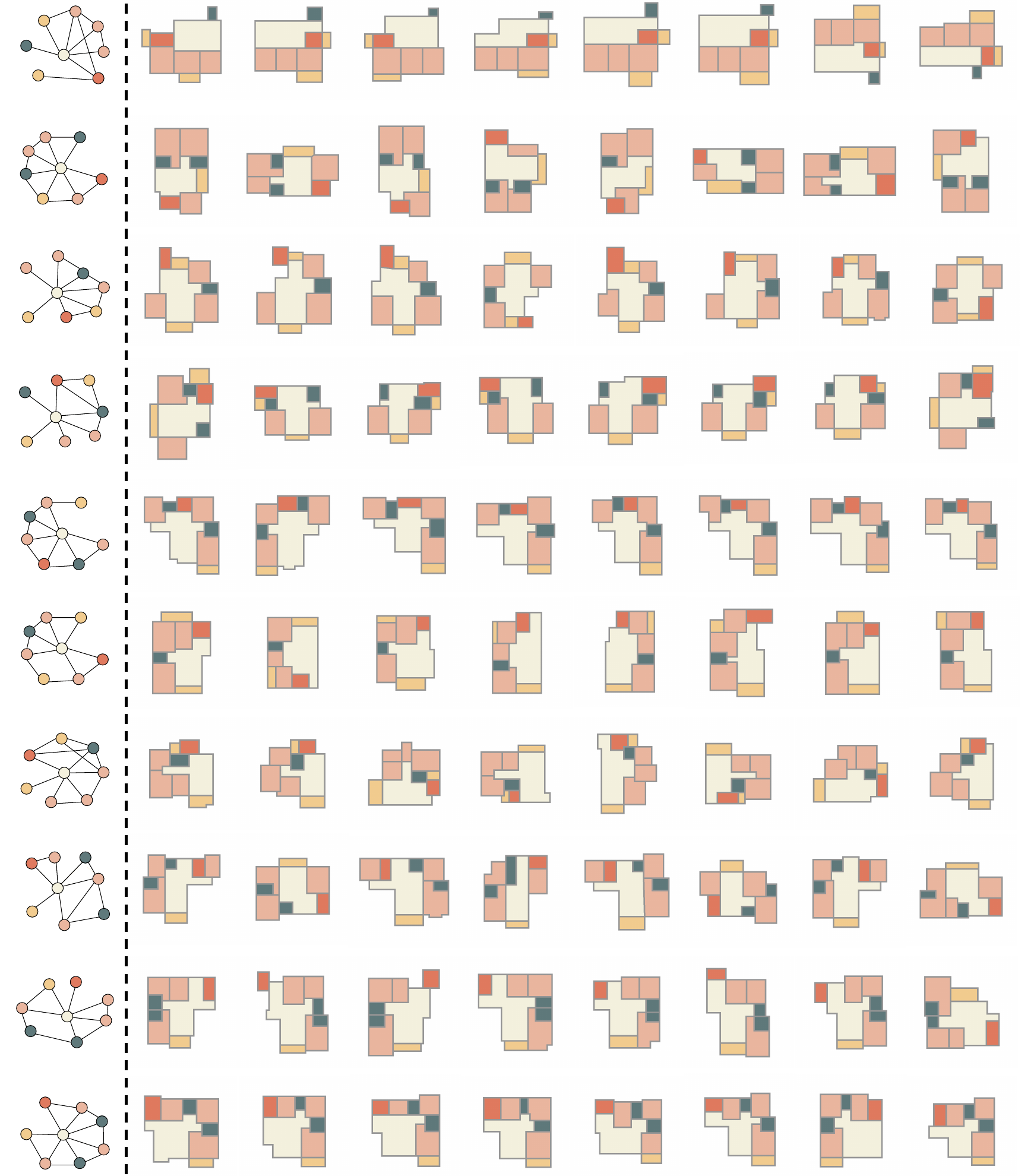}
	\centering
	\caption{More results of topology-constrained generation by our method: part (II).}
	\label{fig:results8}
  \vspace{0mm}
\end{figure*}

Figure~\ref{fig:results1} displays a comparison of boundary-constrained generation across different methods, including the ground-truth (GT), \textit{Graph2Plan}~\cite{hu2020graph2plan}, \textit{WallPlan}~\cite{sun2022wallplan}, and ours. Figure~\ref{fig:results2} provides a comparison of topology-constrained generation among various techniques. including \textit{HouseDiffsion}~\cite{shabani2023housediffusion}, \textit{House-GAN++}~\cite{houseganpp} and ours. 
Figure~\ref{fig:results3} showcases the results of unconstrained generation. Figure~\ref{fig:results4} showcases the results of unconstrained generation with slanted walls. 
Figures~\ref{fig:results5} and~\ref{fig:results6} present the results of boundary-constrained generation by our method, while Figures~\ref{fig:results7} and~\ref{fig:results8} illustrate the results of topology-constrained generation by our method.

\begin{figure*}[t]
	\includegraphics[width=\linewidth]{fig/fig13}
	\centering
	\caption{More results of boundary-constrained generation by our method: part (I).}
	\label{fig:results5}
  \vspace{0mm}
\end{figure*}

\begin{figure*}[t]
	\includegraphics[width=\linewidth]{fig/fig14}
	\centering
	\caption{More results of boundary-constrained generation by our method: part (II).}
	\label{fig:results6}
  \vspace{0mm}
\end{figure*}

\bibliography{aaai25}